\documentclass[12pt,twocolumn]{aastex631}

\usepackage{graphicx}
\usepackage{verbatim}

\newcommand{\n}{\nodata}
\newcommand{\be}{\begin{itemize}}
\newcommand{\ee}{\end{itemize}}
\newcommand{\muasyr}{\hbox{$\; \mu{\rm as \ y}^{-1}\;$}}

\newlength\mystoreparindent

\def\fermi{\textit{Fermi }}

\received{2021 July 19}
\accepted{2021 August 27}
\revised{2021 August 23}
\submitjournal{Astrophysical Journal}

\shorttitle{MOJAVE. XVIII. AGN Jet Kinematics}

\shortauthors{M. L. Lister et al.}

\begin{document}

\title{MOJAVE: XVIII. Kinematics and Inner Jet Evolution of Bright
  Radio-Loud Active Galaxies}

\author[0000-0003-1315-3412]{M. L. Lister}
\affiliation{Department of Physics and Astronomy, Purdue University, 525 Northwestern Avenue,
West Lafayette, IN 47907, USA
}
\author[0000-0002-4431-0890]{D. C. Homan}
\affiliation{Department of Physics, Denison University, Granville, OH 43023, USA}

\author[0000-0002-0093-4917]{K. I. Kellermann}
\affiliation{National Radio Astronomy Observatory, 520 Edgemont Road,
Charlottesville, VA 22903, USA}

\author[0000-0001-9303-3263]{Y. Y. Kovalev}
\affiliation{Astro Space Center of Lebedev Physical Institute,
Profsoyuznaya 84/32, 117997 Moscow, Russia}
\affiliation{Moscow Institute of Physics and Technology,
  Institutsky per. 9, Dolgoprudny, Moscow region, 141700, Russia}
\affiliation{Max-Planck-Institut f\"ur Radioastronomie, Auf dem H\"ugel 69,
53121 Bonn, Germany}

\author[0000-0002-9702-2307]{A. B. Pushkarev}
\affiliation{Crimean Astrophysical Observatory, 98409 Nauchny, Crimea, Russia}
\affiliation{Astro Space Center of Lebedev Physical Institute,
Profsoyuznaya 84/32, 117997 Moscow, Russia}
\affiliation{Moscow Institute of Physics and Technology,
  Institutsky per. 9, Dolgoprudny, Moscow region, 141700, Russia}

\author[0000-0001-9503-4892]{E. Ros}
\affiliation{Max-Planck-Institut f\"ur Radioastronomie, Auf dem H\"ugel 69,
53121 Bonn, Germany}

\author[0000-0001-6214-1085]{T. Savolainen}
\affiliation{Aalto University Department of Electronics and
  Nanoengineering, PL 15500, FI-00076 Aalto, Finland}
\affiliation{Aalto University Mets\"ahovi Radio Observatory, Mets\"ahovintie 114,
FI-02540 Kylm\"al\"a, Finland}
\affiliation{Max-Planck-Institut f\"ur Radioastronomie, Auf dem H\"ugel 69,
53121 Bonn, Germany}

\begin{abstract}
  We have analyzed the parsec-scale jet kinematics of 447 bright
  radio-loud AGN, based on 15 GHz VLBA data obtained between 1994
  August 31 and 2019 August 4. We present new total intensity and
  linear polarization maps obtained between 2017 January 1 to 2019
  August 4 for 143 of these AGN.  We tracked 1923 bright features for
  five or more epochs in 419 jets.  A majority (60\%) of the
  well-sampled jet features show either accelerated
  or non-radial motion. In 47 jets there is at least one
  non-accelerating feature with an unusually slow apparent speed. Most
  of the jets show variations of $10^\circ$ to $50^\circ$ in their
  inner jet position angle (PA) over time, although the overall
  distribution has a continuous tail out to $200^\circ$. AGN with SEDs
  peaked at lower frequencies tend to have more variable PAs, with 
  BL\,Lacs being less variable than quasars. The \fermi LAT gamma-ray
  associated AGN also tend to have more variable PAs than the non-LAT AGN
  in our sample. We attribute these trends to smaller viewing angles 
  for the lower spectral peaked and LAT-associated jets. 
  We identified 13 AGN where multiple features emerge over
  decade-long periods at systematically increasing or decreasing PAs.
  Since the ejected features do not fill the entire jet cross-section,
  this behavior is indicative of a precessing flow instability near
  the jet base. Although some jets show indications of oscillatory PA
  evolution, we claim no bona fide cases of periodicity since
  the fitted periods are comparable to the total VLBA time coverage.

\end{abstract}


\section{INTRODUCTION} 

Relativistic jets from active galactic nuclei (AGN) represent some of
the most energetic long term phenomena in the universe, and played a
key role in regulating galaxy formation via feedback processes
\citep[e.g.,][]{2012ARAA..50..455F}. A powerful tool for investigating these
outflows is the Very Long Baseline Array (VLBA), which provides full
polarization, sub-milliarcsecond scale imaging at radio wavelengths.
The latter has revealed important details of the pc-scale structural
and magnetic field evolution of AGN jets
\citep{2019ARAA..57..467B, 2013EPJWC..6106001W}, and is a critical
driver for state of the art numerical jet simulations
\citep{2020ARA&A..58..407D,2021NewAR..9201610K}.

Since the VLBA's commissioning in 1994, we have carried out a program
to investigate the parsec-scale properties of several hundred of the
brightest AGN jets in the northern sky above J2000 declination
$-30^\circ$. This effort started out as the 2~cm VLBA survey
\citep{1998AJ....115.1295K}, and continued as the MOJAVE survey in
2002 with the addition of full polarization imaging of a complete flux
density-limited sample \citep{MOJAVE_I}. We have presented our
findings in a number of papers in this series, including our most
recent analysis of jet kinematics based on multi-epoch data obtained
between 1994 August 31 and 2016 December 26 \citep{MOJAVE_XVII}.

In this paper we perform a new kinematics analysis that adds VLBA data
taken up to and including 2019 August 4, and increases the number of
AGN jets in our study from 409 to 447.  Some of these jets have
no new data after the cutoff date of our previous kinematics paper
\citep{MOJAVE_XVII}, but we have made some revisions to their model
fits.  For this reason we tabulate here the fit values for all 447
jets, which supersede those presented in previous papers.  We
have excluded several AGN that were in the \cite{MOJAVE_XVII} analysis
since we subsequently determined that they have uncertain core
locations and less reliable kinematic fits. These consist of NGC
  1052, which has a heavily absorbed core at 15 GHz
  \citep{2003AA...401..113V}, five compact symmetric objects
(B2~0026+34, S4~0108+38, S4~0646+60, B3~0710+439, and TXS~2021+614),
and five AGN with extremely compact radio structure that appear
unresolved or barely resolved(PKS 0414$-$189, S5~0615+82,
TXS~0640+090, TXS~1739+522, and B2~2023+33). We will present a
kinematic analysis of the compact symmetric object jets in a future
paper in this series.

The size and time coverage of our data set provides a unique opportunity to examine the stability of the innermost regions of AGN jets over time. Although smoking gun evidence of AGN jet precession has been seen on kpc-scales (e.g., \citealt{2019MNRAS.490.1363S,2010ApJ...713L..74F,1982ApJ...262..478G}), 
there have been few large systematic pc-scale studies to date. In \cite{MOJAVE_X} we identified several 
individual cases of AGN that may be undergoing periodic changes in their inner jet position angle (PA). 
We revisit our earlier study by analyzing  a much larger data set consisting of  173 jets that have 12 or more VLBA observations
acquired over a minimum 10 year period. We provide evidence that over time, AGN eject narrow jet features at different position angles within a broader outflow, resulting in apparent changes in their inner jet direction on the sky. The range and variance of these changes is larger for jets oriented closer to the line of sight.  In some AGN jets the PAs of successively ejected features follow a systematic trend over time, indicating a wobbling flow instability near the jet base from which the features emerge. Occasionally this instability may be disrupted and/or a new instability forms at another location, resulting in a sudden jump in the inner jet PA. 

The layout of our paper is as follows. In Section~\ref{data} we
describe our VLBA data and reduction methods, as well as the general
properties of our AGN jet sample. We describe our kinematics analysis
method in Section~\ref{analysis}, and present our study of inner jet
position angle variations.  We summarize our findings in
Section~\ref{conclusions}.  Throughout this paper we
use the cosmological parameters
$\Omega_m = 0.27$, $\Omega_\Lambda = 0.73$ and $H_o = 71 \;
\mathrm{km\; s^{-1} \; Mpc^{-1}}$ \citep{Komatsu09} and define sky position angles in degrees east of north.

\section{\label{data}OBSERVATIONAL DATA}

\subsection{MOJAVE Data Archive}

 \begin{deluxetable*}{llclcccccc} 
 \tablecolumns{10} 
 \colnumbers
 \tabletypesize{\scriptsize} 
 \tablewidth{0pt}  
 \tablecaption{\label{gentable} MOJAVE AGN Properties}  
 \tablehead{ \colhead{B1950}  & \colhead {Alias} & \colhead{Opt.} & \colhead{$z$}& \colhead{Optical Reference} & \colhead{LAT Association} & \colhead{TeV} & \colhead{1.5 Jy} & \colhead{$\nu_\mathrm{pk}$} & \colhead{Ref.}} 
 \startdata 
 0003+380 &  \objectname{S4 0003+38} & Q& 0.229 & \cite{1994AAS..103..349S} & 4FGL J0005.9+3824 & \n & \n & 13.1 & 10  \\ 
 0003$-$066 &  \objectname{NRAO 005} & B& 0.347 & \cite{2005PASA...22..277J} & 4FGL J0006.3$-$0620 & \n & Y & 12.9 & 9  \\ 
 0006+061 &  \objectname{TXS 0006+061} & B& \n & \cite{2012AA...538A..26R} & 4FGL J0009.1+0628 & \n & \n & 13.4 & 10  \\ 
 0007+106 &  \objectname{III Zw 2} & G& 0.089 & \cite{1970ApJ...160..405S} & \tablenotemark{a} & \n & Y & 13.3 & 1  \\ 
 0011+189 &  \objectname{RGB J0013+191} & B& 0.477 & \cite{2013ApJ...764..135S} & 4FGL J0014.1+1910 & \n & \n & 13.7 & 9  \\ 
 0010+405 &  \objectname{4C +40.01} & Q& 0.256 & \cite{1992ApJS...81....1T} & 4FGL J0013.6+4051 & \n & \n & 12.8 & 9  \\ 
 0012+610 &  \objectname{4C +60.01} & U& \n & \n & 4FGL J0014.8+6118 & \n & \n & 13.1 & 10  \\ 
 0014+813 &  \objectname{S5 0014+813} & Q& 3.382 & \cite{1987AZh....64..262V} & \n & \n & \n & 12.5 & 1  \\ 
 0015$-$054 &  \objectname{PMN J0017-0512} & Q& 0.226 & \cite{2012ApJ...748...49S} & 4FGL J0017.5$-$0514 & \n & \n & 13.6 & 10  \\ 
 0016+731 &  \objectname{S5 0016+73} & Q& 1.781 & \cite{1986AJ.....91..494L} & 4FGL J0019.6+7327 & \n & Y & 12.3 & 9  \\ 
 0019+058 &  \objectname{PKS 0019+058} & B& \n & \cite{2017ApJS..233....3T} & 4FGL J0022.5+0608 & \n & \n & 13.1 & 10  \\ 
\enddata 
\tablecomments{Columns are as follows: 
(1) B1950 name, 
(2) other name,
(3) optical spectroscopic classification, where B = BL Lac, Q = quasar, G = radio galaxy, N = narrow-line Seyfert 1, and U = unknown, 
(4) redshift,
(5) reference for redshift and/or optical classification.
(6) high confidence GeV gamma-ray association from Fermi LAT catalogs, where 4FGL = \cite{4FGL,4FGL-DR2}, 3FGL = \cite{3FGL}, 2FGL = \cite{2FGL}, 1FGL = \cite{1FGL}. 
(7) known TeV gamma-ray emitter (\url{http://tevcat.uchicago.edu}).
(8) member of the VLBA 15 GHz flux density-limited 1.5 Jy Quarter Century Sample \citep{MOJAVE_XVII}.
(9) log frequency of synchrotron peak in spectral energy distribution in Hz.
(10) reference for synchrotron peak frequency, where 
1. {ASDC SED builder (\citealt{2011arXiv1103.0749S}) }
2. {\cite{2011ApJ...740...98M}}
3. {\cite{2008AA...488..867N} }
4. {\cite{2011ApJ...743..171A}}
5. {\cite{2006AA...445..441N} }
6. {\cite{2009ApJ...707...55A}}
7. {\cite{2009ApJ...707L.142A}}
8. {\cite{2015AA...578A..69H} }
9. {\cite{4LAC}}
10. {\cite{2015ApJ...810...14A}}
11. {\cite{2015MNRAS.450.3568X}}
12. {\cite{2017AA...598A..17C} }
13. {\cite{2017ApJS..232...18A}}
14. {\cite{3HSP}}
\\  \\ This table is published in its entirety in the machine-readable format.
      A portion is shown here for guidance regarding its form and content.
}

\tablenotetext{a}{Fermi-LAT detection reported by \cite{2018AA...616A..20A}} 
\tablenotetext{b}{Fermi-LAT detection announced in https://www.astronomerstelegram.org/?read=14383.} 
\end{deluxetable*}

As of 2021, the MOJAVE data
archive\footnote{\url{http://www.physics.purdue.edu/MOJAVE}} consists
of nearly ten thousand 15 GHz (2 cm) VLBA observations of over 500 AGN
dating back to 1994, obtained as part of the 2 cm VLBA survey
\citep{1998AJ....115.1295K}, the NRAO
archive,\footnote{\url{http://archive.nrao.edu}} and the MOJAVE
program.  The MOJAVE data archive provides public access to calibrated
visibility and image FITS files for these observations.  Over time, we
have added AGN on the basis of their correlated flux density,
high-energy gamma-ray emission, or membership in other AGN monitoring
programs.  The minimum criteria are a J2000 declination $>-30\arcdeg$
to ensure sufficient interferometric visibility plane coverage, and a
15~GHz VLBA flux density  larger than $\sim 50$~mJy to
ensure direct fringe detection and the ability to self-calibrate the
data.

In \cite{MOJAVE_XVII}, we compiled a complete flux density limited
AGN sample, the 1.5 Jy Quarter Century MOJAVE sample (1.5JyQC), consisting of
all 232 AGN north of J2000 declination $-30\arcdeg$ known to have
exceeded 1.5 Jy in 15 GHz VLBA flux density at any time between 1994.0
and 2019.0. This is the largest and most complete radio-loud blazar
sample to date, covering 75\% of the entire sky. Being a multi-epoch sample selected only
on the basis of parsec-scale jet emission, it is well-suited for
investigating the effects of relativistic beaming on observed blazar
luminosity functions \citep{MOJAVE_IV}, and the properties of the
misaligned (parent) population \citep{MOJAVE_XVII}.

In \cite{MOJAVE_I, MOJAVE_V,MOJAVE_X,MOJAVE_XV}, we have published 15
GHz total intensity and linear polarization maps from the MOJAVE
program up to 2016 Dec 26.  Here we present new 15 GHz VLBA maps of
143 AGN obtained between 2017 January 3 and 2019 August 4.  The 49 new
AGN included here are members of either the Robopol optical
polarization survey \citep{2021MNRAS.501.3715B}, the \fermi 2FHL catalog
\citep{2FHL}, the LAT monitored list of flaring
sources\footnote{\url{https://heasarc.gsfc.nasa.gov/W3Browse/fermi/fermilasp.html}},
or the MOJAVE-\fermi hard spectrum gamma-ray sample \citep{MOJAVE_XV}.
We list the general properties of all the MOJAVE AGN in
\autoref{gentable}. We have compiled synchrotron peak frequencies from
the literature, or used the fit routines provided by the ASDC spectral
energy distribution (SED) builder \citep{2011arXiv1103.0749S}. The
optical classifications and redshifts are compiled from the NED
database, with the literature references listed in column 5 of
\autoref{gentable}.  Of the 68 AGN with unknown redshift, 57 are
classified as BL~Lacs due to their near-featureless optical spectrum
in the reference listed in \autoref{gentable}. Some of the remaining
11 AGN have been classified as BL~Lacs by other authors on the basis
of their gamma-ray properties, however, we classify these as unknown
since they lack published optical spectra. We provide notes on
selected individual AGN in the \autoref{s:source_notes}.
 
\subsection{Data Reduction}

We processed the VLBA data using standard reduction methods in AIPS
\citep{AIPS}, and self-calibrated and imaged the visibilities in
DIFMAP \citep{DIFMAP}.  We determined the absolute EVPA correction at
each epoch via measurements of stable downstream polarized features,
as described in our previous papers \citep{MOJAVE_XV, MOJAVE_XVII}. The EVPAs were originally anchored using near-simultaneous single dish polarization measurements at 15 GHz made at the U. Michigan Radio Observatory \citep{UMRAO}. 

While comparing our total cleaned VLBA flux densities ($S_\mathrm{VLBA}$)
to near-simultaneous single-dish 15 GHz observations made by the Owens
Valley 40m radio monitoring program ($S_\mathrm{OVRO}$; \citealt{2011ApJS..194...29R}), we
discovered a persistent error affecting the VLBA visibility amplitudes. All VLBA data obtained after 2019 April 15 have suffered from
a systematic $\sim 15\% - 25\%$ decrease in correlated flux density on
all baselines. At the time of writing the cause of this drop is
unknown and is being investigated. We applied flux density correction factors to all the visibilities in four affected VLBA
epochs listed in \autoref{VLBAcorrectionfactors}. We determined these corrections by
first estimating the amount of arcsecond-scale flux density resolved
out by the VLBA ($S_\mathrm{res}$) for individual MOJAVE AGN based on multiple previous
near-simultaneous OVRO--VLBA measurements dating back to 2008.  These
were typically $\lesssim 5$\% of the OVRO flux density, reflecting the
highly core-dominated nature of the MOJAVE sample AGN. At each VLBA
epoch, we determined a correction factor for each AGN $c = (S_\mathrm{OVRO} - S_\mathrm{res}) / S_\mathrm{VLBA}$ and computed the median of these values.

\begin{deluxetable}{llc}
\label{VLBAcorrectionfactors}
\tablecolumns{3}
\tabletypesize{\scriptsize} 
 \tablewidth{0pt}  
 \tablecaption{VLBA Flux Density Correction Factors}  
 \tablehead{\colhead{Epoch} & \colhead{Obs. Code} & \colhead{Correction Factor}}
 \startdata
 2019 June 13 & BL229AX  & 1.18 \\
 2019 June 29 & BL229AY  & 1.15 \\
 2019 July 19 & BL229AZ  & 1.27 \\
 2019 August 4 & BL273A & 1.26 \\
 \enddata
 \end{deluxetable}
 
\subsection{VLBA 15 GHz Maps}

\begin{figure}
\centering
\includegraphics[width=\linewidth]{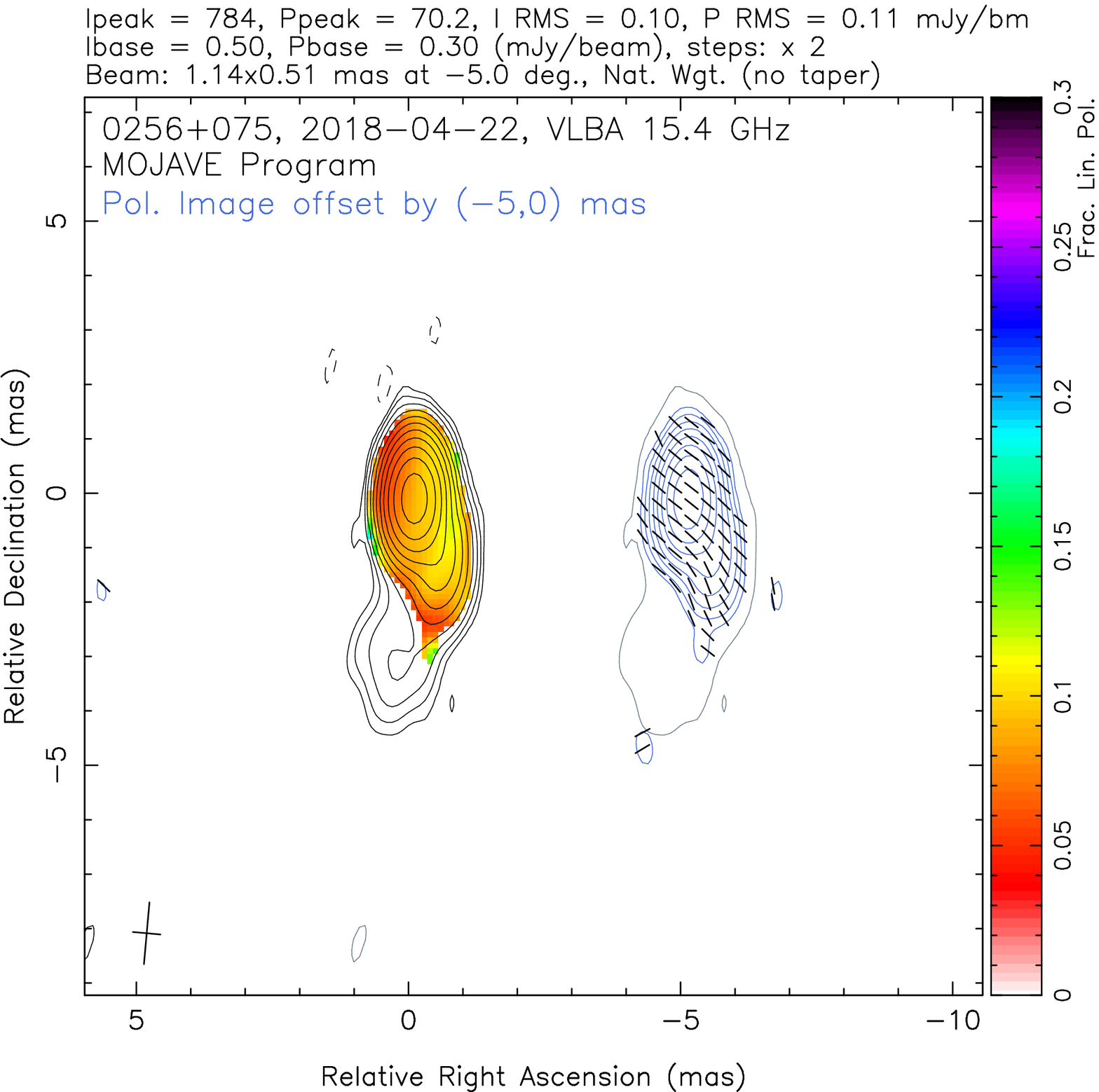}
\caption{\label{mapfigures} 15 GHz VLBA maps of the
  MOJAVE AGN sample. Each panel contains two contour maps of the radio
  source, the first consisting of I contours in successive integer
  powers of two times the lowest contour level, with linear fractional
  polarization overlaid according to the color wedge. A single
  negative I contour equal in magnitude to the base contour level is also plotted with dashed lines.
  The second map includes the lowest positive I contour from the first
  map in black, and linearly polarized intensity contours in blue, also
  increasing by factors of two. The sticks indicate the electric
  polarization vector directions, uncorrected for Faraday rotation.
  The FWHM dimensions and orientation of the elliptical Gaussian
  restoring beam are indicated by a cross in the lower left corner of
  the map. The parameters of each map are listed in \autoref{maptable}. (The complete figure set is available in the online
  journal.) }
\end{figure}

In Figure Set~\ref{mapfigures} we present 648 maps of 142 AGN spanning
epochs from 2017 January 1 to 2019 August 4.  To supplement our
kinematics analysis we processed two data sets (3C~264 and IVS
B1147+245; epoch 2018 March 30, VLBA observation code BM482) from the
VLBA archive, and one from \cite{2020ApJ...899..141L} (TXS 0128+554
epoch 2018 June 29, VLBA observation code BL251).  The remaining data
come from the MOJAVE program (VLBA observation codes BL229 and BL273).

Since any absolute sky positional information is lost during the
self-calibration process, we shifted the origin of each map to the
total intensity Gaussian model-fit position of the core feature, as
described in \S~\ref{modelfitting}. The core is typically the
brightest feature in the map, located at the optically thick surface
close to the base of the jet.

In each 'dual-plot' map in Figure Set~\ref{mapfigures}, we indicate the
FWHM dimensions and orientation of the naturally-weighted elliptical
Gaussian restoring beam by a cross in the lower left corner. The beam
size varies with the declination of the AGN and number of available
antennas, but has typical dimensions of 1.1 mas $\times$ 0.5 mas. We
gridded the Stokes maps with a scale of 0.1 mas per pixel. We
list the parameters of the restoring beam, base contour levels, total
cleaned flux densities, and blank sky map noise levels for each map in
\autoref{maptable}.  The false color corresponds to fractional
polarization, and is superimposed on a total intensity contour map of
the radio emission.  No fractional polarization is plotted in regions
that lie below the lowest total intensity contour level.  The latter
typically corresponds to roughly 3 times the rms noise level of the
map, although this can be higher in the cases of AGN with poorer
interferometric coverage due to extreme southern or near-equatorial
declination, or very bright jet cores due to dynamic range
limitations.

\begin{deluxetable*}{lllcrrcccccrc} 
\tablecolumns{13} 
\colnumbers 
\tabletypesize{\scriptsize} 
\tablewidth{0pt}  
\tablecaption{\label{maptable}15 GHz Map Parameters}  
\tablehead{\colhead{Source} & \colhead {Epoch} & \colhead{$B_\mathrm{maj}$} &\colhead{$B_\mathrm{min}$} & \colhead{$B_\mathrm{pa}$} &  
\colhead{$I_\mathrm{tot}$} &  \colhead{$\sigma_\mathrm{I}$}  &  \colhead{$I_\mathrm{base}$} & 
\colhead{$P_\mathrm{tot}$} & \colhead{$\sigma_\mathrm{Q,U}$}  & \colhead{$P_\mathrm{base}$}&\colhead{EVPA} &  \colhead{Fig.}}
\startdata 
0012+610 & 2017 Jan 3 & 0.80 & 0.64 & $-$6 & 257 & 0.12 & 0.33 & $<$ 0.4 & 0.13 & 0.44 & \n  & 1.1   \\ 
 & 2017 Mar 11 & 0.73 & 0.66 & 0 & 252 & 0.09 & 0.26 & 2.3 & 0.10 & 0.32 & $159$  & 1.2   \\ 
 & 2017 Jun 17 & 0.66 & 0.60 & $-$17 & 232 & 0.10 & 0.23 & 1.6 & 0.11 & 0.30 & $167$  & 1.3   \\ 
 & 2017 Aug 25 & 0.71 & 0.58 & $-$6 & 235 & 0.11 & 0.25 & 2.7 & 0.12 & 0.35 & $177$  & 1.4   \\ 
 & 2018 May 31 & 0.68 & 0.60 & $-$2 & 247 & 0.10 & 0.25 & 2.4 & 0.10 & 0.40 & $162$  & 1.5   \\ 
 & 2019 Jun 29 & 0.86 & 0.66 & $-$28 & 274 & 0.14 & 0.35 & 1.5 & 0.15 & 0.46 & $178$  & 1.6   \\ 
0014+813 & 2017 Jan 28 & 0.71 & 0.66 & 70 & 828 & 0.10 & 0.33 & 10 & 0.10 & 0.35 & $168$  & 1.7   \\ 
 & 2017 Jun 17 & 0.67 & 0.59 & 82 & 853 & 0.10 & 0.30 & 10 & 0.10 & 0.35 & $171$  & 1.8   \\ 
 & 2017 Nov 18 & 0.82 & 0.72 & $-$52 & 922 & 0.11 & 0.90 & 7.1 & 0.13 & 0.51 & $171$  & 1.9   \\ 
 & 2018 Feb 2 & 0.72 & 0.68 & 23 & 904 & 0.11 & 0.34 & 11 & 0.14 & 0.52 & $178$  & 1.10   \\ 
 & 2018 Jul 8 & 0.63 & 0.55 & $-$12 & 849 & 0.12 & 0.35 & 20 & 0.12 & 0.46 & $175$  & 1.11   \\ 
 & 2018 Nov 11 & 0.67 & 0.59 & 42 & 963 & 0.13 & 0.38 & 27 & 0.13 & 0.60 & $177$  & 1.12   \\ 
 & 2019 Jan 19 & 0.63 & 0.61 & 19 & 957 & 0.12 & 0.47 & 26 & 0.18 & 0.65 & $177$  & 1.13   \\ 
 & 2019 Jun 29 & 0.74 & 0.71 & 14 & 1077 & 0.14 & 1.04 & 27 & 0.15 & 0.75 & $2$  & 1.14   \\ 
\enddata 
\tablenotetext{a}{NRAO archive epoch}
\tablecomments{Columns are as follows: (1) B1950 name, (2) date of VLBA observation, (3) FWHM major axis of restoring beam (milliarcseconds), (4) FWHM minor axis of restoring beam (milliarcseconds), (5) position angle of major axis of restoring beam (degrees), (6) total cleaned I flux density (mJy),  (7) rms noise level of Stokes I image (mJy per beam), (8) lowest I contour level (mJy per beam), (9) total cleaned P flux density (mJy), or upper limit, based on 3 times the P rms noise level,  (10) average of blank sky rms noise level in Stokes Q and U images (mJy per beam), (11) lowest linear polarization contour level (mJy per beam), (12) integrated electric vector position angle (degrees), (13) figure number.
\ \ This table is published in its entirety in the machine-readable format.
      A portion is shown here for guidance regarding its form and content.
}
\end{deluxetable*}


We plot a second map of the source, in blue linear polarization
contours increasing by successive factors of two, at an arbitrary sky
position offset from the total intensity map, along with a single
lowest total intensity contour in grey. The overlaid sticks indicate
the observed electric vector directions, and are of arbitrary fixed
length.  We have not corrected their orientations for any Faraday
rotation either internal or external to the AGN jet.  Our rotation
measure study of the MOJAVE sample \citep{MOJAVE_VIII} showed that the
emission from most of these jets experiences only a few degrees of
Faraday rotation at 15 GHz, typically in the region near the base of the jet. The lowest polarization contour is typically 3 times
$\sigma_\mathrm{Q,U}$, but in $\sim 14\%$ of the epochs is more than 5 times
$\sigma_\mathrm{Q,U}$ due to a high peak total intensity in the map,
or residual polarization feed leakage errors. We have not applied any Rician
de-biasing corrections (i.e., $P_\mathrm{corr} = P_\mathrm{obs}\sqrt{1
  - (\sigma_P / P_\mathrm{obs}^2})$; \citealt{1974ApJ...194..249W} ) to the maps, since these are
$\lesssim$ 5\% for regions above our lowest polarization contour
level.

\section{\label{analysis}DATA ANALYSIS}

For our kinematics analysis we have used 15.4 GHz VLBA observations of
447 AGN obtained between 1994 August 31 and 2019 August 4 as part of
the MOJAVE and 2 cm VLBA survey programs, with supplementary data from
the NRAO archive.  There are 49 AGN which have not appeared previously
in any MOJAVE kinematics analysis.  Although the density and span of
time coverage varies considerably among the sample, all of the
AGN in the kinematics analysis have at least 5 high quality VLBA
epochs over a 1.5 year to 25 year time span. 

\subsection{\label{modelfitting}Gaussian Modeling}

We modeled the sky brightness distribution for each VLBA observation
in the $(u,v)$ visibility plane using the {\it modelfit} task in
DIFMAP. We list the properties of the fitted features in
\autoref{gaussiantablestub}.  In some instances, it was impossible to
robustly cross-identify the same features in a jet from one epoch to
the next. We indicate the features with robust cross-identifications
across at least five epochs in column 10 of
\autoref{gaussiantablestub}. In making robustness determinations,
  we considered the consistency of evolution in the sky position, flux
  density, and brightness temperature of the features over time. For
the non-robust features, we caution that the assignment of the same
identification number across epochs does not necessarily indicate a
reliable cross-identification. We initially assigned the
  identification numbers in ascending order roughly based on their
  distance from the core, but in many cases the original order had to
  be modified during the cross-identification process.

\begin{deluxetable*}{lclcrrcrcc} 
\tablecolumns{10} 
\tabletypesize{\scriptsize} 
\tablewidth{0pt}  
\tablecaption{\label{gaussiantablestub}Fitted Jet Features}  
\tablehead{\colhead{} & \colhead {} &   \colhead {} & 
 \colhead{I} & \colhead{r} &\colhead{P.A.} & \colhead{Maj.} & 
\colhead{} &\colhead{Maj. P.A.}   \\  
\colhead{Source} & \colhead {I.D.} &  \colhead {Epoch} & 
\colhead{(mJy)} & \colhead{(mas)} &\colhead{(\arcdeg)} & \colhead{(mas)} & 
\colhead{Ratio} &\colhead{(\arcdeg)}&\colhead{Robust?}   \\  
\colhead{(1)} & \colhead{(2)} & \colhead{(3)} & \colhead{(4)} &  
\colhead{(5)} & \colhead{(6)} & \colhead{(7)} & \colhead{(8)} & 
 \colhead{(9)} &  \colhead{(10)}} 
\startdata 
0003+380  & 0& 2006 Mar 9  & 489  & 0.04 & 290.7 & 0.23 & 0.33 & 292 & Y\\ 
0003+380  & 1& 2006 Mar 9  & 7.2  & 3.98 & 121.8 & 0.72 & 1 & \n & Y\\ 
0003+380  & 2& 2006 Mar 9  & 42.1  & 1.25 & 110.5 & 0.51 & 1 & \n & Y\\ 
0003+380  & 6& 2006 Mar 9  & 104  & 0.28 & 114.6 & 0.27 & 1 & \n & Y\\ 
0003+380  & 7& 2006 Mar 9  & 2.9  & 2.31 & 119.3 & \n & \n & \n & N\\ 
0003+380  & 0& 2006 Dec 1  & 320  & 0.10 & 308.1 & 0.25 & 0.29 & 295 & Y\\ 
0003+380  & 1& 2006 Dec 1  & 4.8  & 3.65 & 120.8 & 1.63 & 1 & \n & Y\\ 
0003+380  & 2& 2006 Dec 1  & 20.9  & 1.56 & 111.0 & 0.25 & 1 & \n & Y\\ 
0003+380  & 5& 2006 Dec 1  & 22.9  & 0.75 & 116.2 & 0.32 & 1 & \n & Y\\ 
0003+380  & 6& 2006 Dec 1  & 145  & 0.45 & 116.3 & 0.05 & 1 & \n & Y\\ 
\enddata 

\tablenotetext{a}{Individual feature epoch not used in kinematic fits.}

\tablecomments{Columns are as follows: (1) B1950 name, (2) feature identification number (zero indicates core feature), (3) observation epoch, (4) flux density at 15 GHz in mJy, (5) position offset from the core feature (or map center for the core feature entries) in milliarcseconds, (6) position angle with respect to the core feature (or map center for the core feature entries) in degrees,  (7) FWHM major axis of fitted Gaussian in milliarcseconds, (8) axial ratio of fitted Gaussian, (9) major axis position angle of fitted Gaussian in degrees, (10) robust feature flag. 
\ \ This table is published in its entirety in the machine-readable format.
      A portion is shown here for guidance regarding its form and content.}
\end{deluxetable*}

Based on previous analysis \citep{MOJAVE_VI}, we estimate the typical
uncertainties in the feature centroid positions to be $\sim 20$\% of
the FWHM naturally-weighted image restoring beam dimensions. For
isolated bright and compact features, the positional errors are
smaller by approximately a factor of two.  We estimate the formal
errors on the feature sizes to be roughly twice the positional error,
according to \cite{1999ASPC..180..301F}.  The flux density accuracies
are approximately 5\% (see \autoref{s:source_notes} of
\citealt{2002ApJ...568...99H}), but can be significantly larger for
features located very close to one another. Also, at some epochs which
lacked data from one or more antennas, the fit errors of some features
are much larger.  We do not use the latter in our kinematics or jet
position angle analyses, and indicate them with flags in
\autoref{gaussiantablestub}.

\subsection{Jet Kinematics}

We analyzed the kinematics of all individual robust jet features using
three methods: (i) a simple one-dimensional radial motion fit, (ii) a
non-accelerating vector fit in two dimensions (right ascension and
declination), and (iii) a constant acceleration two-dimensional fit
(for features with ten or more epochs).  We use the radial fit for
diagnostic purposes only, and do not tabulate those fit results here.
In all cases, we assume the bright core feature (id~0 in
\autoref{gaussiantablestub}) to be stationary, and measure the
positions of jet features at all epochs with respect to it. We
described the details of the fitting method in \cite{MOJAVE_XVII}.  We
note that flaring activity, a not-yet-resolved newly-ejected feature,
or the variable core-shift effect \citep{2019MNRAS.485.1822P} can
sometimes result in core positional variations. In cases where a large
shift of the core was observed, we flagged that epoch from the
kinematics analysis.

We made radial and vector motion fits using all of the available data
from 1994 August 31 to 2019 August 4 on 1923 robust jet features in
447 jets. A total of 28 of these jets had no robust
features for kinematic analysis due to weak/absent downstream jet flux density, an insufficiently stable core feature, and/or insufficient angular resolution.

In \autoref{vectormotiontablestub} we list the results of our vector
motion fits. Due to the nature of our kinematic model, which naturally
includes the possibility of accelerated motion, we did not estimate
ejection epochs (Column 12) for any features where we could not
confidently extrapolate their motion to the core.  Jet features for
which we list an ejection epoch had the following properties: (i)
significant motion $(\mu \ge 3\sigma_\mu)$, (ii) no significant
acceleration, (iii) a velocity vector direction $\phi$ within
$15^\circ$ of the outward radial direction to high confidence, i.e.,
$|\langle\vartheta\rangle - \phi|+2\sigma \le 15^\circ$, where
$\vartheta$ is the mean position angle, (iv) an extrapolated position
at the ejection epoch no more than $0.2$ mas from the core, and (v) a
fitted ejection epoch that differed by no more than 0.5 years from
that given by the radial motion fit.

Approximately half ($N = 926$) of the robust features met the $\ge 10$
epoch criterion for an acceleration fit, and we tabulate these results
in \autoref{accelmotiontablestub}.  The majority (60\%) of these
well-sampled features display either significant acceleration or
non-radial motion.

\begin{figure*}
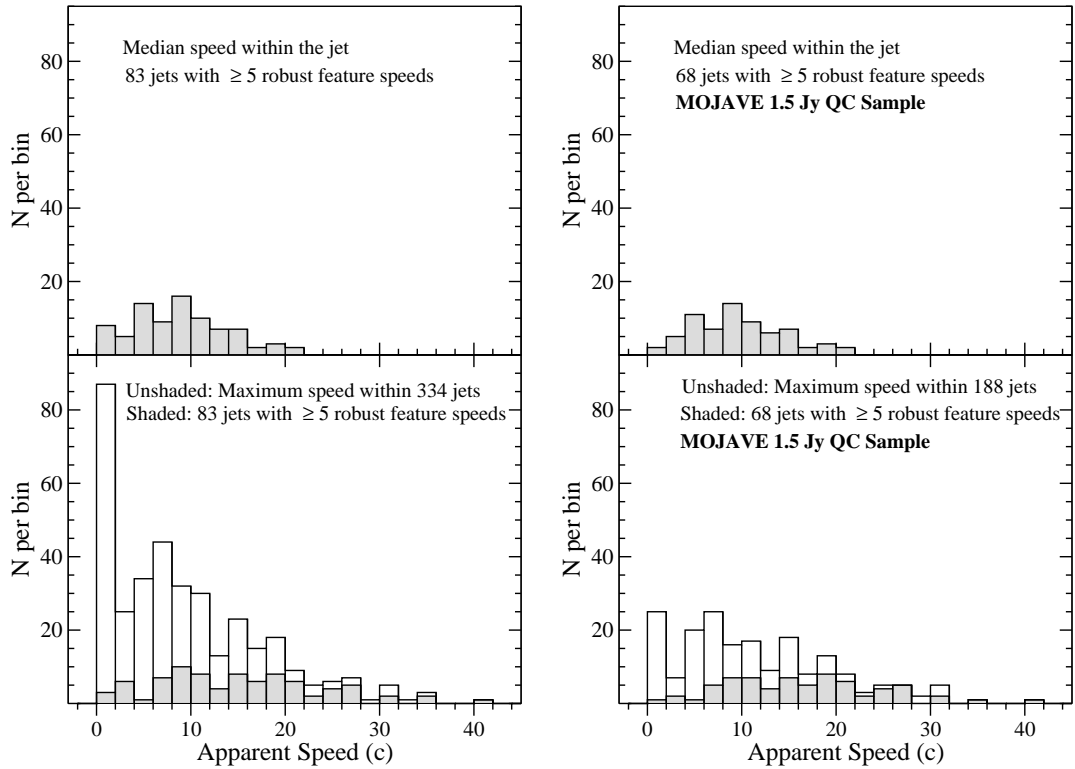

\centering
\includegraphics[width=0.4\textwidth,trim=-0.5cm -0.5cm -0.5cm -0.5cm]{betaspeedhist.eps}
\includegraphics[width=0.4\textwidth,trim=-0.5cm -0.5cm -0.5cm -0.5cm]{1.5JyQC_betaspeedhist.eps}
\caption{\label{betaspeedhist} Histogram plots of median jet speed (top panels)
  and fastest jet speed (bottom panels) for AGN with known
  redshifts. The histograms in the right hand panels show AGN in the flux density-limited MOJAVE 1.5 Jy QC sample.}
\end{figure*}

We have marked 64 features in \autoref{accelmotiontablestub}, in 47
different AGN, that have appreciably slower speeds than other features
in the same jet.  These slow pattern speed features  (i) do
not have a $\ge 3\sigma$ acceleration, (ii) have an angular speed smaller
than 20 \muasyr, and (iii) have a speed at least ten times slower than
the fastest feature in the same jet.

We also include flags in \autoref{vectormotiontablestub} and
\ref{accelmotiontablestub} for inward-moving features. In
\cite{MOJAVE_XVII} we discuss the possibility that some of these may
be the result of centroid shifts in diffuse emission regions, or
curved trajectories crossing the line of sight. Of the 1923 robust
features in our latest analysis, only 48 show apparent inward motion.
Although this fraction (2.5\%) is small, 39 of the AGN jets in our
sample (8.7\%) show this phenomenon. We provide updated
information on selected AGN with inward-moving features in \autoref{s:source_notes}. 
In an earlier paper \citep{MOJAVE_V} we discussed the five AGN in the
original flux density-limited MOJAVE sample that have counter-jet
features: 0238$-$084 (NGC 1052), 0316+413 (3C 84), 1228+126 (M87), PKS
1413+135, and 1957+405 (Cyg A). With the exception of PKS 1413+135,
these are all nearby ($z < 0.06$) radio galaxies with jets much closer
to the plane of the sky than the other AGN in our sample. PKS 1413+135
is a peculiar AGN with blazar properties that may be a
gravitationally lensed system (see \citealt{2021ApJ...907...61R} and
references therein). In our current analysis we find only two new
examples of candidate counter-jets in the 1.5JyQC sample: 1928+738 and
1253$-$055 (3C 279), but in both cases we believe that these are not
actual counter-jets, and that the true core is visible only at some
epochs (see Appendix notes).  Among the non-1.5JyQC AGN presented in
this paper, there are only a handful with counter-jet features. These
consist of two compact symmetric objects TXS 0128+554 and 1509+054
(PMN J1511+0518), two giant radio galaxies 1637+826 (NGC 6251) and
2043+749 (4C +74.26), the nearby ($z = 0.029$) radio galaxy 0305+039
(3C 78), and the quasar 1148$-$001 (4C $-$00.47). In the latter AGN,
we have assumed the core to be the most compact feature of the jet, but
higher frequency VLBA observations are needed to more precisely
determine its location. In Table~\ref{gaussiantablestub} we also list a
counter-jet feature for the high redshift $z = 2.624$ quasar PKS
B0742+103, but the core location is uncertain in this source as noted
in the Appendix.

\begin{deluxetable*}{lcrrrrrrrrrrrrr} 
\rotate 
\tablecolumns{15} 
\tabletypesize{\scriptsize} 
\tablewidth{0pt}  
\tablecaption{\label{vectormotiontablestub}Vector Motion Fit Properties of Jet Features}  
\tablehead{\colhead{} & \colhead {} &   \colhead {} & 
\colhead{$\langle S\rangle$}  &\colhead{$\langle R\rangle$} &\colhead{$\langle d_{\mathrm{proj}}\rangle$} & \colhead{$\langle\vartheta\rangle$} & 
 \colhead{$\phi$}&   \colhead{$ |\langle\vartheta\rangle - \phi|$}  &\colhead{$\mu$}  & \colhead{$\beta_{app}$} & && \colhead{$\alpha_m$}& \colhead{$\delta_m$}    \\  
\colhead{Source} & \colhead {I.D.} &  \colhead {N} & 
\colhead{(mJy)} &\colhead{(mas)} & \colhead{(pc)} & \colhead{(deg)}   & 
\colhead{(deg)}& \colhead{(deg)} &\colhead{($\mu$as y$^{-1})$}& \colhead{($c$)}  &\colhead{$t_{ej}$}  & \colhead{$t_\mathrm{mid}$}& \colhead{($\mu$as)}& \colhead{($\mu$as)}  \\  
\colhead{(1)} & \colhead{(2)} & \colhead{(3)} & \colhead{(4)} &  
\colhead{(5)} & \colhead{(6)} & \colhead{(7)} & \colhead{(8)} & 
 \colhead{(9)}& \colhead{(10)}&  
\colhead{(11)} & \colhead{(12)} & \colhead{(13)} & \colhead{(14)} & \colhead{(15)}   } 
\startdata 
0003+380  & 1 & 8  & 5 &4.23&  15.36& $ 120.7$ & 96$\pm$17 & 24$\pm$17 & 158$\pm$43 & 2.30$\pm$0.63 & \n &  2008.81 &3691$\pm$74 & $-$2169$\pm$80\\ 
0003+380  & 2 & 6  & 19 &1.78&  6.45& $ 112.6$ & 120.1$\pm$3.1 & 7.5$\pm$3.1 & 317$\pm$25 & 4.61$\pm$0.36 & \n &  2007.71 &1662$\pm$29 & $-$694$\pm$11\\ 
0003+380  & 4 & 5  & 16 &1.25&  4.53& $ 114.9$ & 205$\pm$14 & 90$\pm$14\tablenotemark{b} & 39$\pm$10 & 0.57$\pm$0.15 & \n &  2009.54 &1130$\pm$11 & $-$527$\pm$14\\ 
0003+380  & 5 & 8  & 40 &0.75&  2.71& $ 117.5$ & 21$\pm$89 & 96$\pm$89 & 2.7$\pm$7.6 & 0.04$\pm$0.11 & \n &  2010.26 &663$\pm$20 & $-$342$\pm$10\\ 
0003+380  & 6 & 10  & 98 &0.39&  1.43& $ 115.4$ & 335$\pm$46 & 141$\pm$46 & 12.7$\pm$8.4\tablenotemark{d} & 0.19$\pm$0.12 & \n &  2009.90 &350$\pm$22 & $-$158$\pm$19\\ 
0003$$-$$066  & 2 & 5  & 222 &1.05&  5.12& $ 322.9$ & 226.3$\pm$4.9 & 96.6$\pm$5.0\tablenotemark{b} & 191$\pm$15 & 4.09$\pm$0.33 & \n &  1997.80 &$-$585.9$\pm$8.9 & 883$\pm$37\\ 
0003$$-$$066  & 3 & 9  & 119 &2.82&  13.73& $ 296.9$ & 284.8$\pm$4.7 & 12.1$\pm$4.8 & 250$\pm$39 & 5.36$\pm$0.83 & \n &  1999.33 &$-$2375$\pm$98 & 1237$\pm$41\\ 
0003$$-$$066  & 4\tablenotemark{a} & 26  & 120 &6.61&  32.23& $ 285.6$ & 284$\pm$11 & 2$\pm$11 & 41$\pm$14 & 0.87$\pm$0.29 & \n &  2004.83 &$-$6326$\pm$60 & 1768$\pm$22\\ 
0003$$-$$066  & 5\tablenotemark{a} & 14  & 1031 &0.70&  3.40& $ 10.7$ & 350.9$\pm$5.3 & 19.9$\pm$5.5\tablenotemark{b} & 88.1$\pm$4.3 & 1.888$\pm$0.091 & \n &  2004.37 &138$\pm$18 & 634.1$\pm$9.0\\ 
0003$$-$$066  & 6\tablenotemark{a} & 10  & 97 &1.01&  4.92& $ 290.2$ & 210$\pm$15 & 81$\pm$15\tablenotemark{b} & 55$\pm$17 & 1.18$\pm$0.37 & \n &  2003.78 &$-$941$\pm$15 & 359$\pm$33\\ 
\enddata

\tablenotetext{a}{Acceleration model fit indicates significant accelerated motion.}
\tablenotetext{b}{Feature has significant non-radial motion according to the vector motion fit.}
\tablenotetext{c}{Feature has significant inward motion according to the vector motion fit.}
\tablenotetext{d}{Feature has slow pattern speed.}

\tablenotetext{\phn}{A question mark indicates a feature whose motion is not consistent with outward, radial motion but for which the possibility of inward motion and its degree of non-radialness are uncertain.}

\tablecomments{Columns are as follows: (1) B1950 name, (2) feature number, (3) number of fitted epochs, (4) mean flux density at 15 GHz in mJy,  (5) mean distance from core feature in mas, (6) mean projected distance from core feature in pc, (7) mean position angle with respect to the core feature in degrees, (8) position angle of velocity vector in degrees, (9) offset between mean position angle and velocity vector position angle in degrees,  (10) proper motion in $\mu$as y$^{-1}$, (11) apparent speed in units of the speed of light, (12) estimated epoch of origin, (13) date of reference (middle) epoch used for fit,  (14) fitted right ascension position with respect to the core at the middle epoch in $\mu$as, (15) fitted declination  position with respect to the core at the middle epoch in $\mu$as.
\\ \\This table is published in its entirety in the machine-readable format.
      A portion is shown here for guidance regarding its form and content.}
\end{deluxetable*} 
\begin{deluxetable*}{lcrrrrrrrrrr} 
\tablecolumns{12} 
\rotate 
\tabletypesize{\scriptsize} 
\tablewidth{0pt}  
\tablecaption{\label{accelmotiontablestub}Acceleration Fit Properties of Jet Features}  
\tablehead{\colhead{} & \colhead {}& \colhead{$\phi$} &    \colhead{$ |\langle\vartheta\rangle - \phi|$}& 
\colhead{$\mu$} & \colhead{$\beta_{app}$} & \colhead{$ \dot{\mu} $}  & \colhead{$ \psi $}  & \colhead{$ \dot{\mu}_\perp $}  & \colhead{$ \dot{\mu}_\parallel $}  &  \colhead{$\alpha_\mathrm{m}$} &\colhead{$\delta_\mathrm{m}$} \\  
\colhead{Source} & \colhead {I.D.} & \colhead{(deg)} & \colhead{(deg)} &   
\colhead{($\mu$as y$^{-1})$}& \colhead{(c)} &\colhead{($\mu$as y$^{-2})$}&\colhead{(deg)}& \colhead{($\mu$as y$^{-2})$}& \colhead{($\mu$as y$^{-2})$}  & \colhead{($\mu$as)}& \colhead{($\mu$as)}   \\  
\colhead{(1)} & \colhead{(2)} & \colhead{(3)} & \colhead{(4)} &  
\colhead{(5)} & \colhead{(6)} & \colhead{(7)}  & \colhead{(8)} & \colhead{(9)}  & \colhead{(10)}& \colhead{(11)}  & \colhead{(12)}  } 
\startdata 
0003+380  & 6  & $ 333 \pm 44 $ & $ 142 \pm 44 $& $ 13.4 \pm 8.6 $ & 0.20$\pm$0.12 &  $ 9.8 \pm 8.4$ & $ 309 \pm 53$ & $ -4.0 \pm 9.4$ & $ 9.0 \pm 9.0$     & $371\pm 33$ & $ -175 \pm 28$\\ 
0003$-$066  & 4\tablenotemark{a}  & $ 277.3 \pm 3.8 $ & $ 8.3 \pm 3.8 $& $ 50.9 \pm 5.3 $ & 1.09$\pm$0.11 &  $ 28.5 \pm 2.3$ & $ 73.7 \pm 3.1$ & $ 11.4 \pm 2.1$ & $ -26.1 \pm 2.5$     & $-6582\pm 32$ & $ 1693 \pm 20$\\ 
0003$-$066  & 5\tablenotemark{a}  & $ 353.9 \pm 3.0 $ & $ 16.8 \pm 3.1\tablenotemark{b} $& $ 87.2 \pm 4.4 $ & 1.868$\pm$0.093 &  $ 26.6 \pm 4.9$ & $ 274 \pm 10$ & $ -26.3 \pm 4.9$ & $ 4.5 \pm 4.8$     & $199\pm 15$ & $ 630 \pm 14$\\ 
0003$-$066  & 6\tablenotemark{a}  & $ 211.3 \pm 9.6 $ & $ 78.9 \pm 9.6\tablenotemark{b} $& $ 54 \pm 11 $ & 1.16$\pm$0.24 &  $ 65 \pm 16$ & $ 336 \pm 11$ & $ 54 \pm 13$ & $ -37 \pm 18$     & $-901\pm 16$ & $ 268 \pm 35$\\ 
0003$-$066  & 8\tablenotemark{a}  & $ 290.7 \pm 1.6 $ & $ 3.5 \pm 1.6 $& $ 330.4 \pm 9.7 $ & 7.08$\pm$0.21 &  $ 67 \pm 12$ & $ 127 \pm 10$ & $ -19 \pm 12$ & $ -64 \pm 12$     & $-2444\pm 30$ & $ 1121 \pm 28$\\ 
0003$-$066  & 9  & $ 295.2 \pm 4.1 $ & $ 7.5 \pm 4.3 $& $ 278 \pm 20 $ & 5.96$\pm$0.42 &  $ 99 \pm 35$ & $ 110 \pm 22$ & $ 9 \pm 37$ & $ -99 \pm 35$     & $-1769\pm 52$ & $ 582 \pm 53$\\ 
0010+405  & 1  & $ 340.7 \pm 4.4 $ & $ 11.9 \pm 4.4 $& $ 432 \pm 42 $ & 6.99$\pm$0.68 &  $ 44 \pm 83$ & $ 147 \pm 76$ & $ 11 \pm 53$ & $ -43 \pm 70$     & $-4259\pm 76$ & $ 6991 \pm 107$\\ 
0010+405  & 2  & $ 9 \pm 123 $ & $ 41 \pm 123 $& $ 2 \pm 14 $ & 0.04$\pm$0.23 &  $ 4 \pm 22$ & $ 152 \pm 123$ & $ 2 \pm 21$ & $ -3 \pm 23$     & $-898\pm 30$ & $ 1470 \pm 48$\\ 
0010+405  & 3  & $ 138 \pm 83 $ & $ 170 \pm 83 $& $ 2.5 \pm 5.4 $ & 0.041$\pm$0.088 &  $ 6.9 \pm 6.1$ & $ 99 \pm 57$ & $ -4.3 \pm 8.8$ & $ 5.4 \pm 9.0$     & $-493.6\pm 9.5$ & $ 783 \pm 15$\\ 
0010+405  & 4  & $ 113 \pm 98 $ & $ 145 \pm 98 $& $ 1.4 \pm 4.5 $ & 0.022$\pm$0.072 &  $ 5.2 \pm 8.9$ & $ 318 \pm 69$ & $ -2.2 \pm 7.1$ & $ -4.7 \pm 8.1$     & $-240.6\pm 9.0$ & $ 382 \pm 14$\\ 
\enddata 
\tablenotetext{a}{Feature shows significant accelerated motion.}
\tablenotetext{b}{Feature shows significant non-radial motion according to the acceleration fit.}
\tablenotetext{c}{Feature shows significant inward motion according to the acceleration fit.}
\tablenotetext{\phn}{A question mark indicates a feature whose motion is not consistent with outward, radial motion but for which the possibility of inward motion and its degree of non-radialness are uncertain.}

\tablecomments{Columns are as follows: (1) B1950 name, (2) feature number, (3) proper motion position angle in degrees,  (4) offset between mean position angle and proper motion position angle in degrees, (5) proper motion in $\mu$as  y$^{-1}$, (6) apparent speed in units of the speed of light, (7) acceleration in $\mu$as  y$^{-2}$, (8) acceleration vector position angle in degrees, (9)  acceleration perpendicular to velocity direction in $\mu$as  y$^{-2}$, (10)  acceleration parallel to velocity direction in $\mu$as  y$^{-2}$, (11) fitted right ascension position with respect to the core at the middle epoch in $\mu$as, (12) fitted declination  position with respect to the core at the middle epoch in $\mu$as. 
\\ \\ This table is published in its entirety in the machine-readable format.
      A portion is shown here for guidance regarding its form and content.}

\end{deluxetable*}

 \begin{deluxetable*}{llcclcccrrcr}
 \tablecolumns{12} 
 \colnumbers
 \tablewidth{0pt}  
 \tablecaption{\label{jettable} MOJAVE Jet Properties}  
 \tablehead{ \colhead{B1950}  & \colhead{$\mu_\mathrm{max}$} &\colhead{$\mu_\mathrm{med}$} & \colhead{$\beta_\mathrm{max}$}  & \colhead{$\beta_\mathrm{med}$} &\colhead{$N_\mathrm{ep}$} &\colhead{$N_\mathrm{r}$} &\colhead{Ref.} &\colhead{$\overline{\mathrm{PA}}$} & \colhead{$\Delta \mathrm{PA}$} & \colhead{log Var(PA)} & \colhead{$\Delta t$}}
 \startdata 
 0003+380 & 317 $\pm$ 25    &   \n  & 4.61 $\pm$ 0.36   & \n &10 & 5 &  3  & 115 & 17 & $ -2.5$ & 7.4 \\ 
 0003$-$066 & 330.4 $\pm$ 9.7    &   116 $\pm$ 23  & 7.08 $\pm$ 0.21   & 2.48 $\pm$ 0.49 &27 & 9 &  3  & 15 & 12 & $ -2.8$ & 17.3 \\ 
 0006+061 & 221 $\pm$ 43    &   \n  & \n   & \n &5 & 2 &  6  & 63 & 4 & $ -3.5$ & 1.4 \\ 
 0007+106 & 269 $\pm$ 50    &   \n  & 1.58 $\pm$ 0.29   & \n &25 & 2 &  3  & 292 & 32 & $ -2.2$ & 12.9 \\ 
 0011+189 & 159 $\pm$ 16    &   \n  & 4.54 $\pm$ 0.46   & \n &8 & 2 &  3  & 219 & 3 & $ -3.9$ & 2.1 \\ 
 0010+405 & 428 $\pm$ 40    &   \n  & 6.92 $\pm$ 0.64   & \n &12 & 4 &  3  & 328 & 4 & $ -4.0$ & 5.2 \\ 
 0012+610 & 7.2 $\pm$ 6.3  \tablenotemark{a}  &   \n  & \n   & \n &6 & 2 &  6  & 37 & 4 & $ -3.5$ & 2.5 \\ 
 0014+813 & 87.8 $\pm$ 8.5    &   \n  & 9.47 $\pm$ 0.91   & \n &14 & 3 &  6  & 184 & 22 & $ -2.1$ & 22.7 \\ 
 0015$-$054 & 50 $\pm$ 20  \tablenotemark{a}  &   \n  & 0.72 $\pm$ 0.28  \tablenotemark{a} & \n &8 & 1 &  3  & 242 & 16 & $ -2.2$ & 3.5 \\ 
 0016+731 & 98.5 $\pm$ 4.1    &   \n  & 7.64 $\pm$ 0.32   & \n &16 & 2 &  6  & 126 & 43 & $ -1.8$ & 24.8 \\ 
 0019+058 & 257 $\pm$ 35    &   \n  & \n   & \n &7 & 2 &  3  & 285 & 14 & $ -2.5$ & 2.6 \\ 
\enddata 
\tablecomments{Columns are as follows: 
(1) B1950 name, 
(2) maximum jet speed in $\mu$as y$^{-1}$,
(3) median jet speed in $\mu$as y$^{-1}$,
(4) maximum jet speed in units of the speed of light,
(5) median jet speed in units of the speed of light,
(6) number of VLBA epochs,
(7) number of robust fitted jet features,
(8) reference for jet kinematics, where
1: \cite{MOJAVE_X},
2: \cite{MOJAVE_XIII},
3: \cite{MOJAVE_XVII},
4: \cite{2010ApJ...723.1150P},
5: \cite{2017ApJ...846...98J},
6: this paper,
(9) mean innermost jet position angle in degrees,
(10) range of innermost jet position angle in degrees,
(11) log of circular variance of innermost jet position angle,
(12) time coverage used to determine jet position angle statistics in years.
\\ \\
This table is published in its entirety in the machine-readable format.
      A portion is shown here for guidance regarding its form and content.
}

\tablenotetext{a}{Speed is $<3 \sigma$}

\end{deluxetable*}


 In \autoref{jettable} we tabulate a median speed for each jet having at least 5 features with  $\ge 3 \sigma$ speeds; we excluded any counter-jet features, and those flagged as having inward motion. For features with $\ge 3 \sigma$ accelerations we used the speed from the acceleration fit, otherwise we used the vector motion fit speed. The error for the median speed in \autoref{jettable} is either that of the middle value in the distribution, or the mean of the two middle values for jets with an even number of speed measurements.  We also tabulated a maximum speed for each jet using the same criteria. If a jet had no $\ge 3 \sigma$ speed features, we used the fastest $\ge 2 \sigma$ speed, or otherwise, the speed with the lowest error value.  

In the left hand panels of \autoref{betaspeedhist} we plot the 
distribution of median and maximum apparent jet speeds of all the AGN in our sample with
known redshifts. We show the distributions for the 1.5 Jy QC sample AGN in the right hand panels. The maximum speed distribution for the full sample (lower left panel) is peaked below $2c$, but 45\% of the jets in the first bin have 2
or fewer robust features. The maximum speed distribution for jets with
at least five robust features (lower panel, shaded) is not sharply peaked, 
suggesting that multiple features must typically be tracked in a jet to get a
more robust measure of the overall flow speed. The 1.5 Jy QC sample AGN have typically been tracked in the MOJAVE program for longer time periods, so their maximum speed distribution is not as sharply peaked. 

\begin{figure*}
\centering
\includegraphics[width=0.9\linewidth]{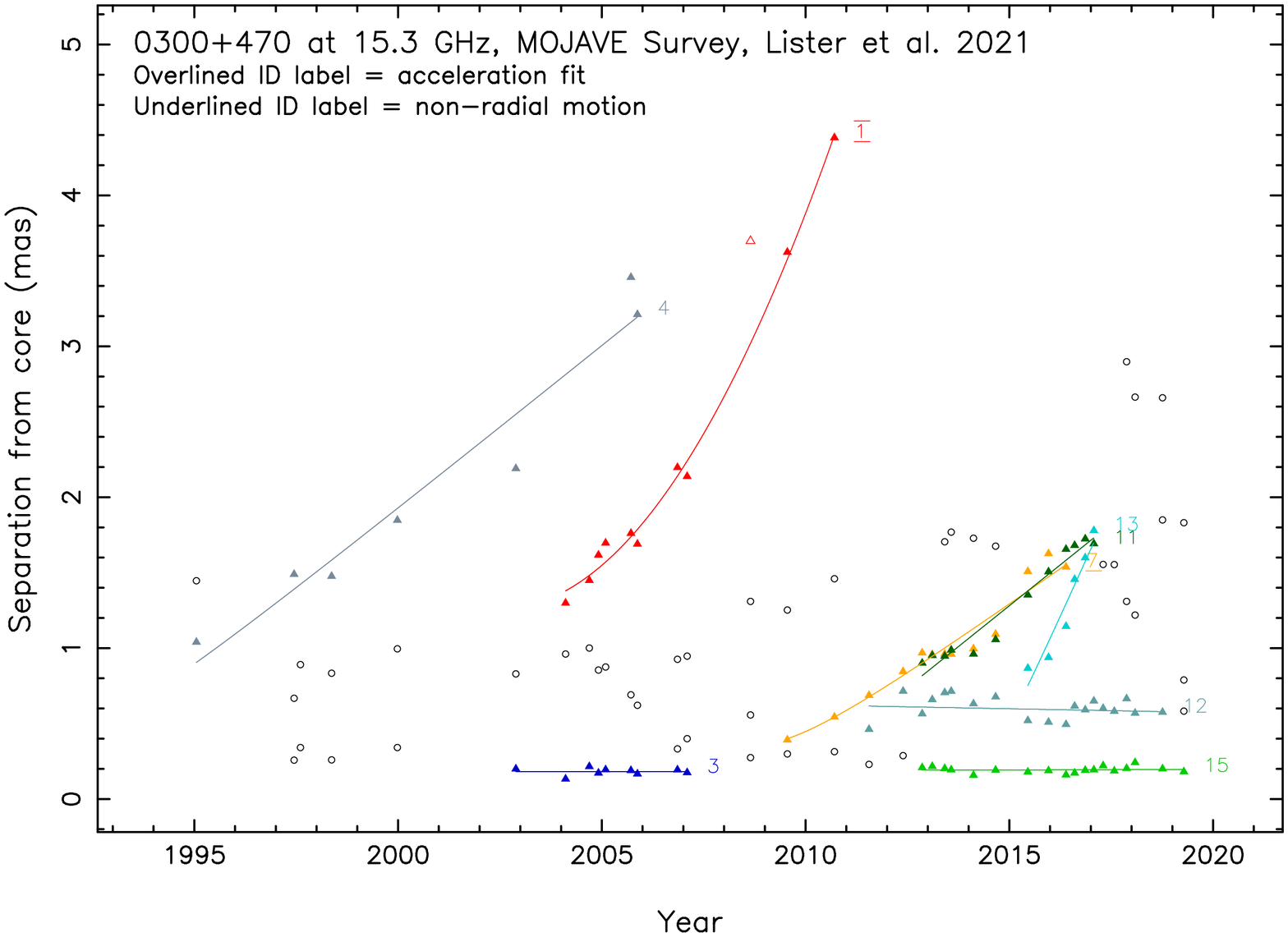}
\caption{\label{sepvstime}
  Plot of angular separation
  from the core versus time for Gaussian jet features.  Colored symbols
  indicate robust features for which kinematic fits were obtained. The
  identification number is overlined if the acceleration model was fit
  and indicated a $\ge 3\sigma$ acceleration. An underlined
  identification number indicates a feature with non-radial motion.
  The $1 \sigma$ positional errors on the individual points typically
  range from 10\% of the FWHM restoring beam dimension for isolated
  compact features, to 20\% of the FWHM for weak features. This
  corresponds to roughly 0.03 mas to 0.15 mas, depending on the
  declination. (The complete figure set is available in the online journal.)}
\end{figure*}

\begin{figure*}
\centering
\includegraphics[angle=270,width=\linewidth]{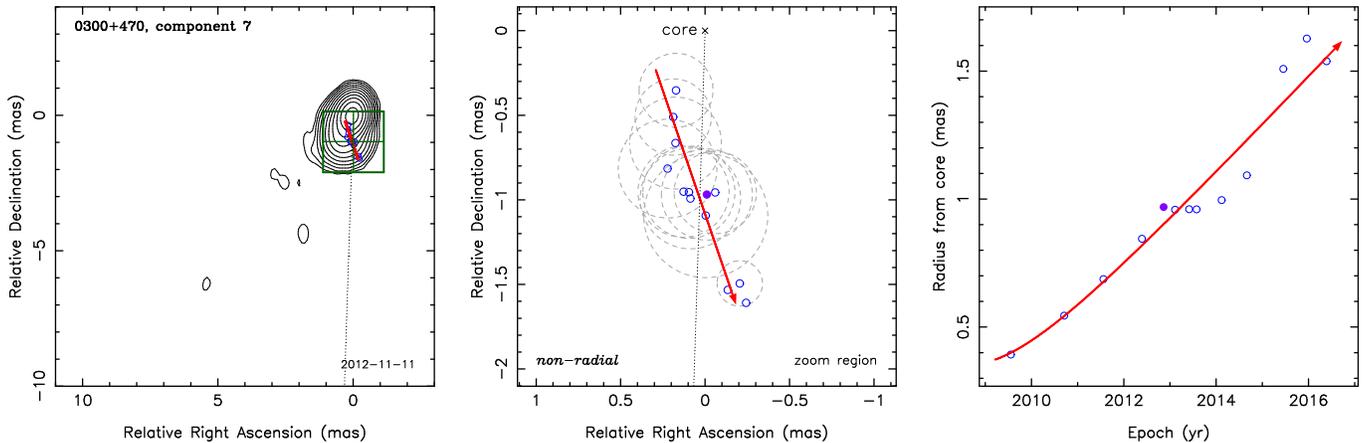}
\caption{\label{xyplot} Motion fits and sky position
  plots of individual robust jet features. Positions are relative to
  the core position.  The left-hand panel shows a 15 GHz VLBA total
  intensity contour image of the jet at the epoch closest to the
  middle reference epoch. The green box delimits the zoomed region
  that is displayed in the middle panel. The feature's position at the
  image epoch is indicated by the green cross-hairs. The dotted line
  connects the feature with the core feature and is plotted with the
  mean position angle.  The position at the image epoch is shown by a
  filled blue circle while other epochs are plotted with unfilled blue
  circles. The red solid line indicates the vector fit (or
  accelerating fit, if there is significant acceleration) to the
  feature positions. The gray dashed circles or ellipses indicate the
  fitted FWHM sizes of the feature at the measured epochs. The right-hand panel shows the radial separation of the feature from the core feature over time. (The complete figure set is available in the online journal.)} 
\end{figure*}

In Figure Set~\ref{sepvstime} we plot the angular separation of
features from the core in each jet versus time. The robust features
have filled colored symbols and solid lines representing
the fit. The feature identification number is overlined if
the acceleration model was fit and yielded a $\ge 3\sigma$ acceleration.
An underlined identification number indicates a feature with
non-radial motion, i.e., its velocity vector does not point back to the
core location within the errors.  We plot the individual trajectories
and fits on the sky for all the robust features in Figure
Set~\ref{xyplot}.

\subsection{Inner Jet Position Angle Analysis}

In \cite{MOJAVE_X} we presented the first large survey of inner
PA variations in AGN jets based on data from 1994 to 2011, where we found evidence
for large changes in jet PA within $\sim 1$ mas ($\sim 8$ pc projected at $z = 1$) of the core over decadal timescales, in some cases as
fast as $10^\circ$ yr$^{-1}$. Here we carry out a new analysis, using
additional data obtained between 2011 May 21 and 2019 August 4.   Our
main goals are to: (i) determine the PA of the jet as close as
possible to the core feature, using the maximum angular resolution of
the VLBA data at 15 GHz, and (ii) examine the behavior of the inner
jet PA over time in individual jets and among different AGN classes.

In our previous study we determined the inner jet PA using a
flux-density weighted position angle average of components from the
\textsc{CLEAN} imaging algorithm within an annular region from 0.15 mas to 1
mas from the core. This method fails, however, for AGN with faint
jet structure and/or insufficient \textsc{CLEAN} components near the core.
Because the method relies on images restored with a Gaussian
(natural-weighted) beam, it also does not yield optimal values for
many jets in the MOJAVE sample that have sharp apparent bent jet 
ridge lines near the core. The latter are better traced using the
Gaussian model components that are modeled in the visibility plane,
which takes advantage of the high positional measurement accuracy of the VLBA for
bright features.

We have therefore chosen to use the position of the innermost Gaussian
model component with respect to the core to measure the jet PA at each
epoch. This generally yields robust results, as exhibited by the
good continuity of jet PAs over time in individual AGN. In a few rare cases there are two Gaussian model components located roughly equidistant from the core, such that the measured PA can alternate between the features over time. 

For those source-epochs where the innermost jet feature is robust, we
used the residuals in the vector motion fit (or
acceleration fit, if a $\ge 3 \sigma$ acceleration) to estimate the PA
measurement error. For each source-epoch we calculated  the standard
deviation of the PA distribution of 10 000 simulated locations scattered
about the innermost feature position. The scatter in R.A. and
declination of these points corresponded to the motion fit residual
$\sigma's$ of the feature in the respective sky directions. 

\begin{figure}
\centering
\includegraphics[width=\linewidth,trim=1.5cm 0.5cm 2.5cm 1.5cm,clip]{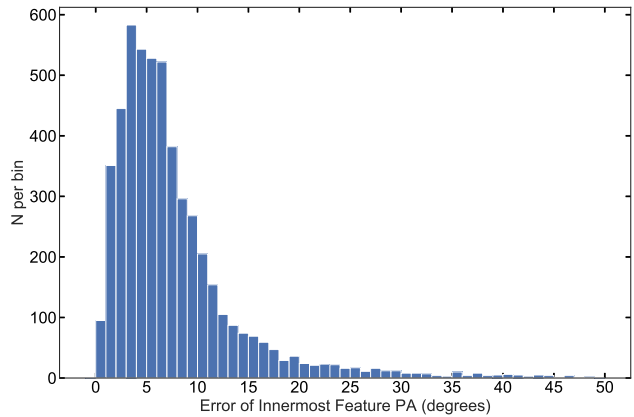}
\caption{\label{dPA_source-epochs_hist} Distribution of innermost jet PA measurement error for all source-epochs. }
\end{figure}

With this method we were able to estimate a PA measurement error for
5167 of the 7562 source-epochs (68\%) for which we tabulated an
innermost jet PA. We plot the distribution of these errors in
\autoref{dPA_source-epochs_hist}. The majority of the values lie
between $1^\circ$ and $10^\circ$, with the most common error being
$\sim 4^\circ$. A small number of source-epochs have large PA error
values. Most of these are overestimates, due to features with
second-order accelerations that are not well-fit by a constant
acceleration model (e.g., PKS 1510$-$089 id 15, 1641+399 id 12), and
therefore have large fit residuals.

\subsubsection{Trends with time}

We calculated for each jet the circular mean ($\overline{PA}$), the
full range over which the PA varies ($\Delta PA$), and the circular
variance. Although the latter quantity formally spans a possible range
between $0 \le$ Var(PA) $\le 1$, we use the base 10 logarithm of Var(PA) as
there are a large number of values very close to zero. We list these
quantities in \autoref{jettable}.

To mitigate the possible effects of sampling bias, we restricted our
statistical analysis to those jets with 12 or more VLBA epochs over a
minimum 10 year period. A total of 173 jets met this
  criterion, 143 of which are in the MOJAVE 1.5JyQC sample.
Most of the jets have inner PAs that vary on decadal timescales over a
range of $10^\circ$ to $50^\circ$.  However some jets display very
large changes in PA, with the overall distribution having a continuous
tail out to $\Delta{PA} = 200^\circ$.

\begin{figure*}
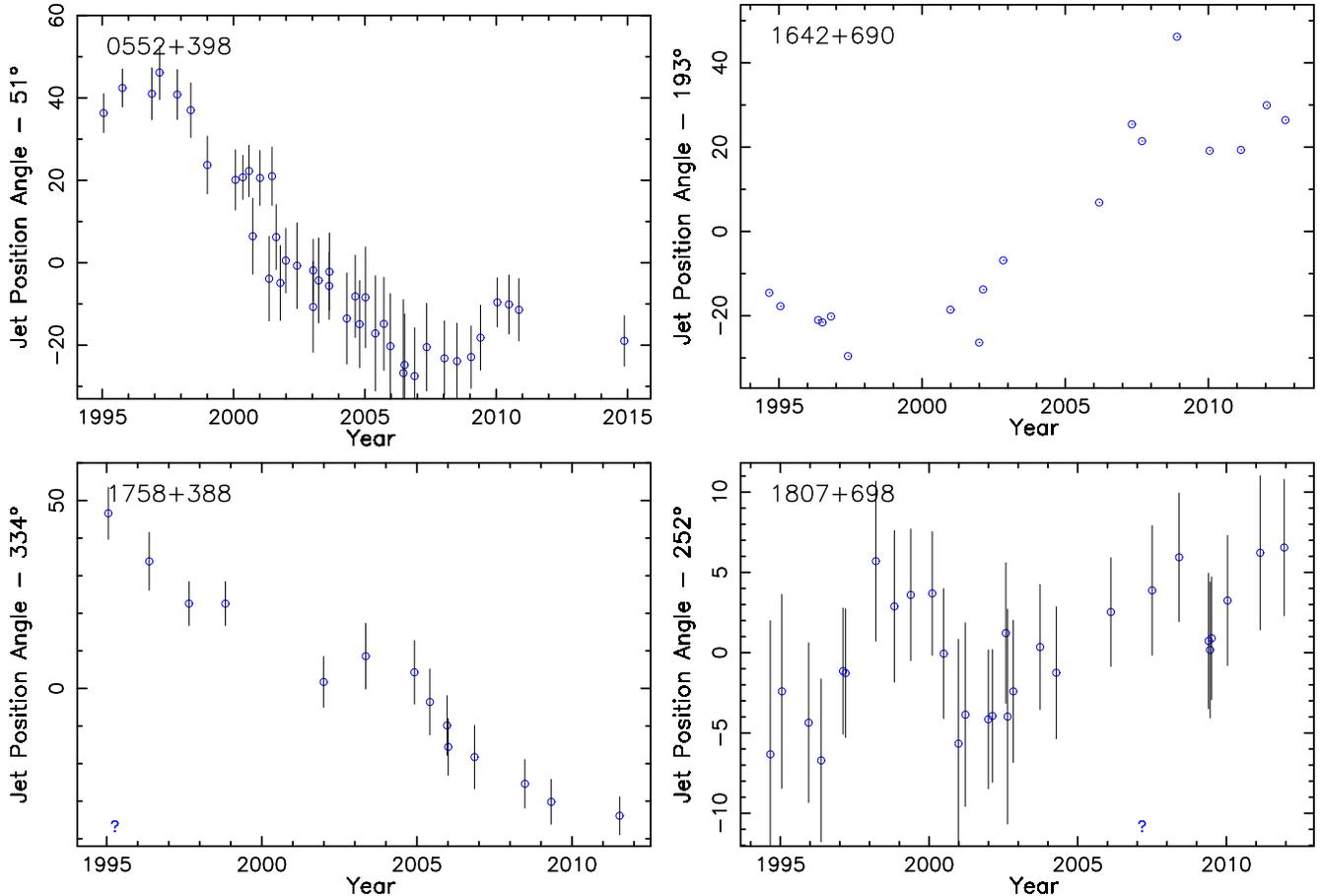

\centering

\includegraphics[angle=270,width=0.49\linewidth,trim=-0.2cm -0.2cm -0.2cm -0.2cm]{0552+398pa.eps}
\includegraphics[angle=270,width=0.49\linewidth,trim=-0.2cm -0.2cm -0.2cm -0.2cm]{1642+690pa.eps}
\includegraphics[angle=270,width=0.49\linewidth,trim=-0.2cm -0.2cm -0.2cm -0.2cm]{1758+388pa.eps}
\includegraphics[angle=270,width=0.49\linewidth,trim=-0.2cm -0.2cm -0.2cm -0.2cm]{1807+698pa.eps}

\caption{\label{PAvstime1}  Plots of inner jet PA versus time for
  selected AGN that show smooth PA evolution due to a
  feature moving steadily outward on a non-radial or curved trajectory. 
  Error bars are plotted for epochs in which the PA error could be estimated from the kinematic fit residuals for the innermost jet feature.
  Question marks indicate VLBA epochs for which a PA measurement could
  not be obtained.}
\end{figure*}

\begin{figure*}
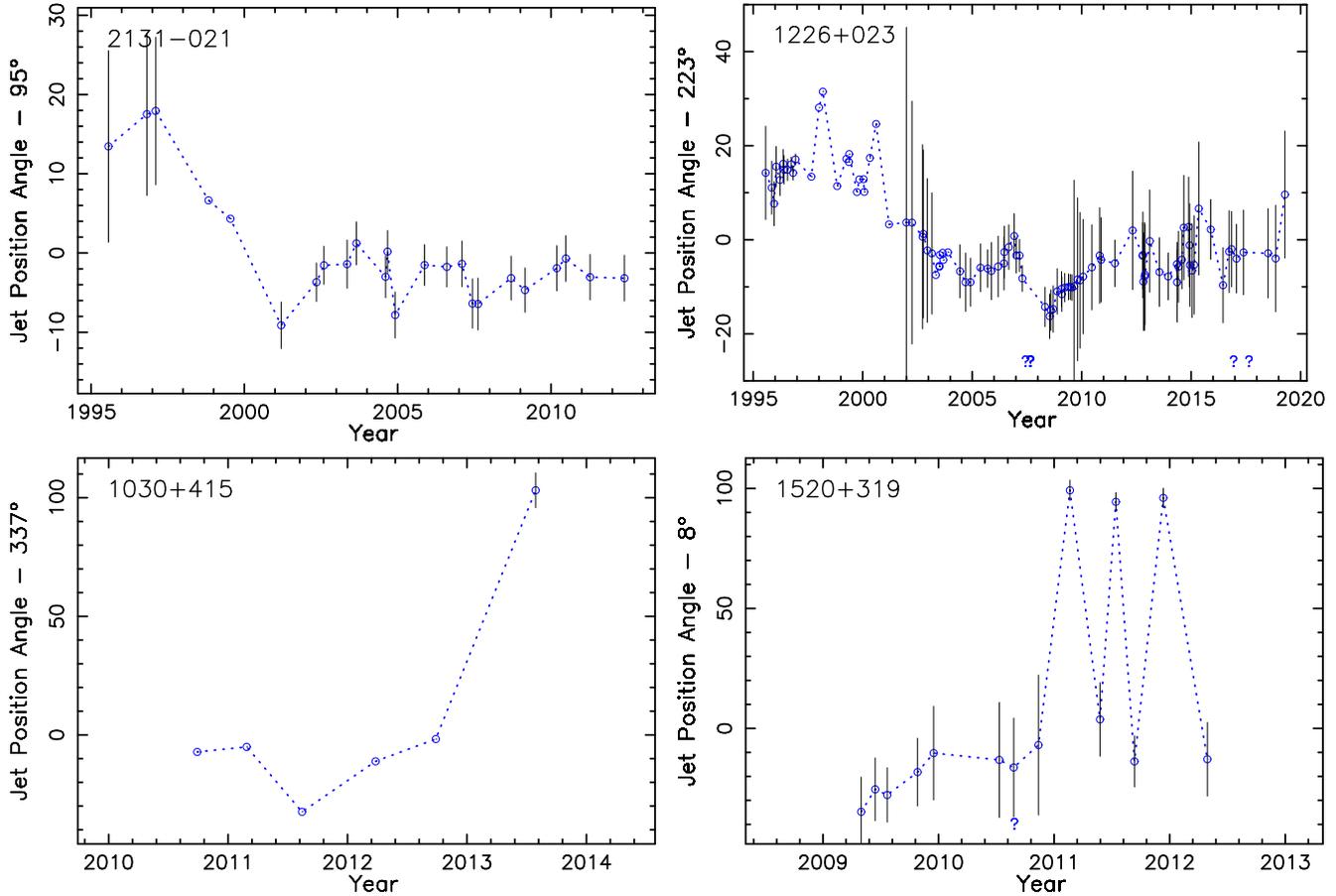

\centering
\includegraphics[angle=270,width=0.49\linewidth,trim=-0.2cm -0.2cm -0.2cm -0.2cm]{2131-021pa.eps}
\includegraphics[angle=270,width=0.49\linewidth,trim=-0.2cm -0.2cm -0.2cm -0.2cm]{1226+023pa.eps}
\includegraphics[angle=270,width=0.49\linewidth,trim=-0.2cm -0.2cm -0.2cm -0.2cm]{1030+415pa.eps}
\includegraphics[angle=270,width=0.49\linewidth,trim=-0.2cm -0.2cm -0.2cm -0.2cm]{1520+319pa.eps}
\caption{\label{PAvstime2} Plots of inner jet PA versus
  time for selected individual AGN.   In \objectname{IVS B2131$-$021} the PA changed in
  2001 due to the emergence of a new jet feature (id~5). The jet of
  1226+023 (\objectname{3C 273}) showed a decade long steady evolution of jet PA
  from 1995 to 2005, while 11 separate features emerged from the core
  region during this time interval.  The jets of \objectname{IVS
    B1030+415} and \objectname{B2 1520+319} contain features with
  position angles differing by $\sim 100^arcdeg$ that are nearly
  equidistant from the core, causing the innermost PA to alternate
  between them at different epochs. }
\end{figure*}

A jet can change its apparent PA over time in several different ways.
One way is to eject a feature that moves steadily outward on a curved
or non-radial trajectory, resulting in a smooth evolution of the inner
jet PA. We show some examples of this in \autoref{PAvstime1}.  A newly
ejected feature may also experience a very rapid fading, reverting the
innermost PA to the next feature farther downstream (e.g.,
\objectname{IVS B2131$-$021}; \autoref{PAvstime2}). In rare cases the
inner jet PA can alternate over time between two nearly equidistant
downstream features (e.g., \objectname{IVS B1030+415} and
\objectname{B2 1520+319}; \autoref{PAvstime2}).

\begin{figure*}
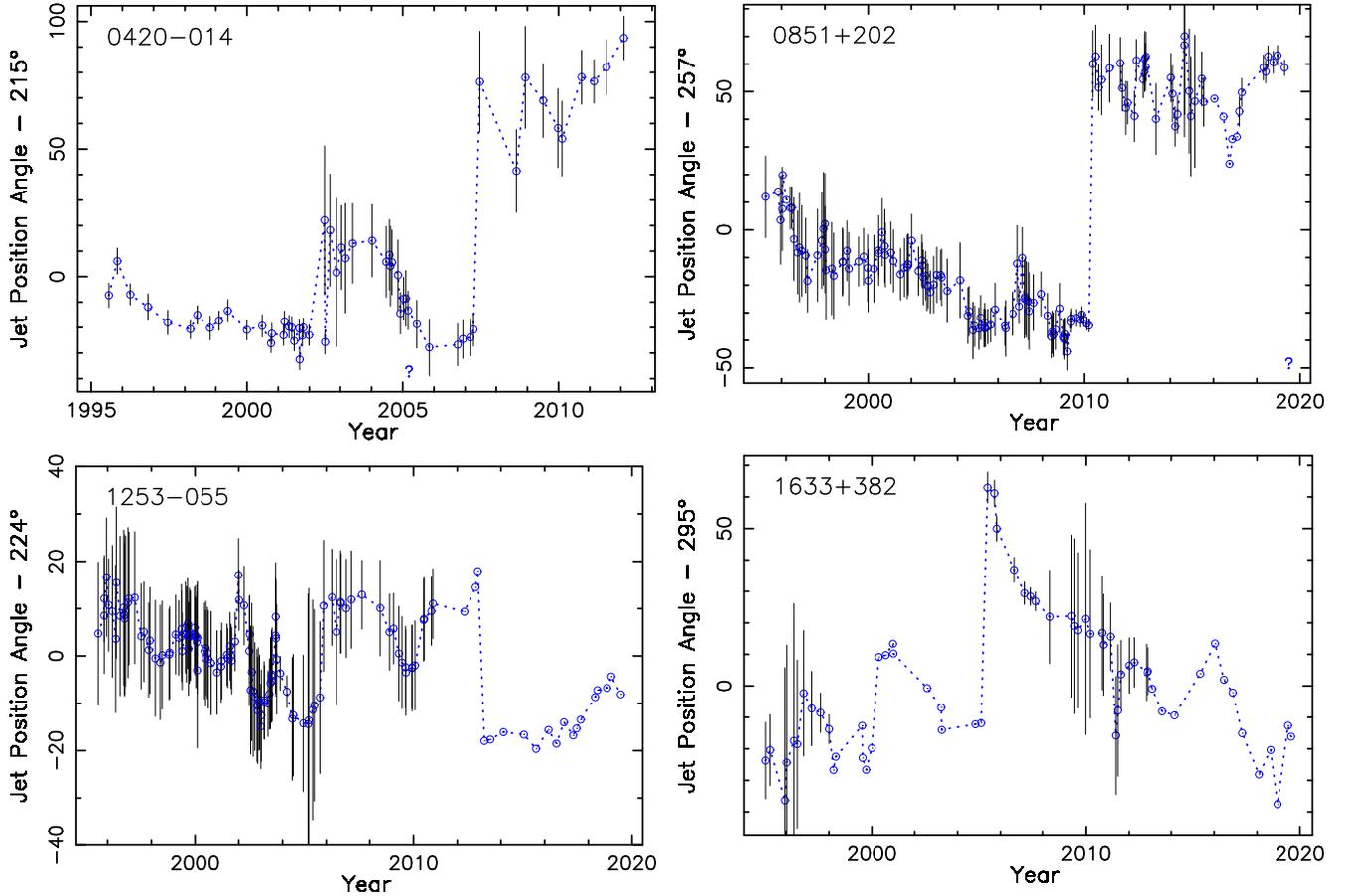

\centering
\includegraphics[angle=270,width=0.49\linewidth,trim=-0.2cm -0.2cm -0.2cm -0.2cm]{0420-014pa.eps}
\includegraphics[angle=270,width=0.49\linewidth,trim=-0.2cm -0.2cm -0.2cm -0.2cm]{0851+202pa.eps}
\includegraphics[angle=270,width=0.49\linewidth,trim=-0.2cm -0.2cm -0.2cm -0.2cm]{1253-055pa.eps}
\includegraphics[angle=270,width=0.49\linewidth,trim=-0.2cm -0.2cm -0.2cm -0.2cm]{1633+382pa.eps}
\caption{\label{PAvstime3}  Plots of inner jet PA versus time for
  several individual AGN that show sudden jumps in PA due to the emergence of
  a new feature with a substantially different trajectory than
  previous features. This occurred in 2002 and 2007 for 0420$-$014, in
2010 for 0851+202 (OJ~287), in 2005 and 2013 for 1253$-$055 (3C~279),
and in 2005 for 1633+382. }
\end{figure*}

It is also possible for a jet to show a sudden large jump in PA when a
new bright feature emerges from the core region at a significantly
different position angle from previous ones.  An example
is 0851+202 (\objectname{OJ 287}), which underwent a monotonic $\sim 3$
degree per year change in jet PA from 1995 to 2010 due to the
non-radial motion of feature id  10.  In May 2010 the inner PA jumped $\sim
100^\circ$ when a new bright feature (id  19) emerged from
the core region with a substantially different trajectory. Other
examples have occurred in the jets of PKS 0420$-$014, 1253$-$055
(\objectname{3C 279}),
and 1633+382 (\objectname{4C +38.41}) (\autoref{PAvstime3}).

\begin{figure*}
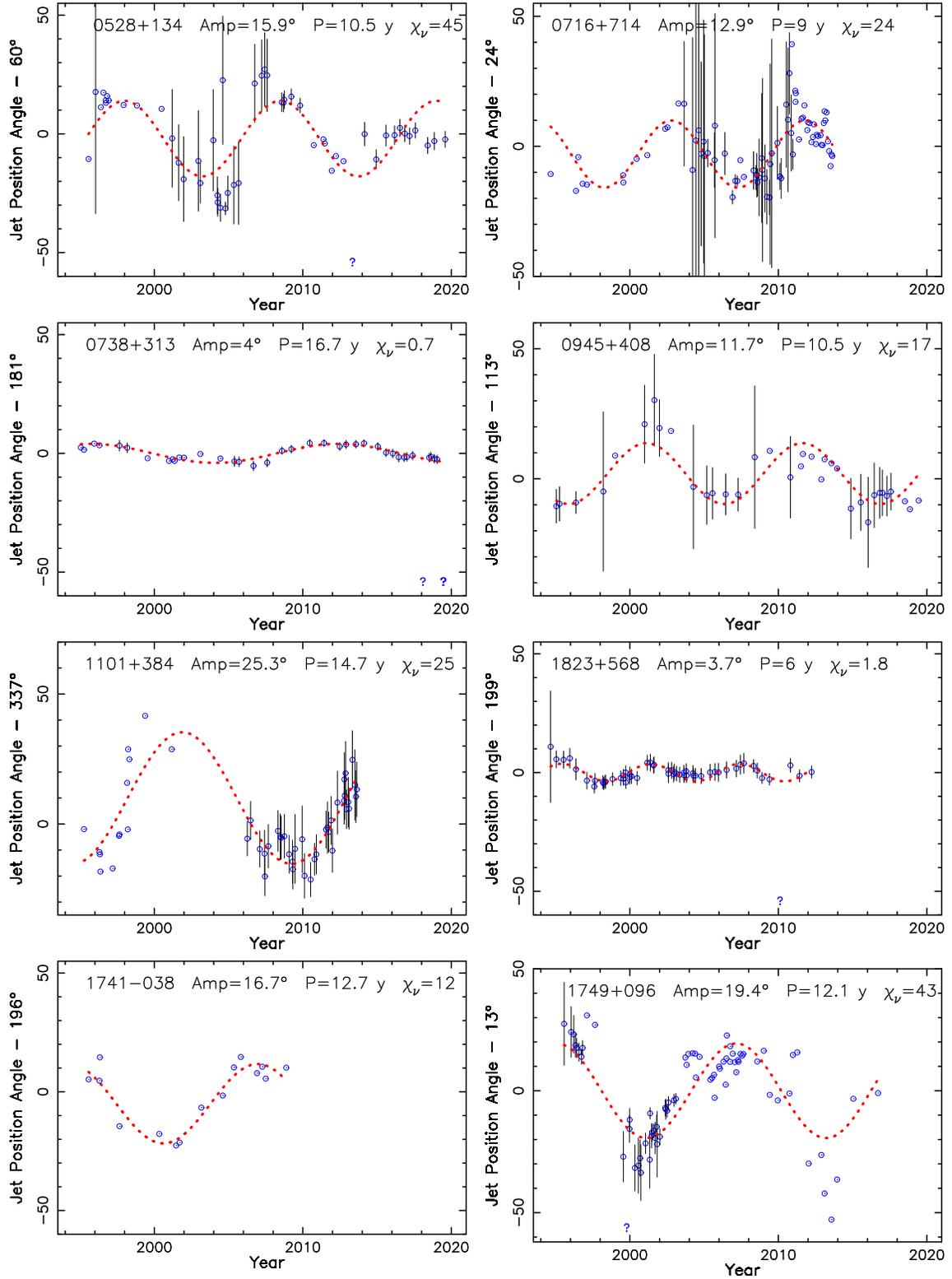

\centering
\includegraphics[angle=270,width=0.42\linewidth,trim=-0.2cm -0.2cm -0.2cm -0.2cm]{0528+134pa.eps}
\includegraphics[angle=270,width=0.42\linewidth,trim=-0.2cm -0.2cm -0.2cm -0.2cm]{0716+714pa.eps}
\includegraphics[angle=270,width=0.42\linewidth,trim=-0.2cm -0.2cm -0.2cm -0.2cm]{0738+313pa.eps}
\includegraphics[angle=270,width=0.42\linewidth,trim=-0.2cm -0.2cm -0.2cm -0.2cm]{0945+408pa.eps}
\includegraphics[angle=270,width=0.42\linewidth,trim=-0.2cm -0.2cm -0.2cm -0.2cm]{1101+384pa.eps}
\includegraphics[angle=270,width=0.42\linewidth,trim=-0.2cm -0.2cm -0.2cm -0.2cm]{1823+568pa.eps}
\includegraphics[angle=270,width=0.42\linewidth,trim=-0.2cm -0.2cm -0.2cm -0.2cm]{1741-038pa.eps}
\includegraphics[angle=270,width=0.42\linewidth,trim=-0.2cm -0.2cm -0.2cm -0.2cm]{1749+096pa.eps}
\caption{\label{sinefits} Plots of inner jet PA
          versus time for individual AGN (blue circles) overplotted
          with best fit sinusoid curves (red dotted lines). These represent the best cases in the sample for possible periodicity. Each panel
          is plotted with the same vertical scaling for comparison
          purposes.  }
\end{figure*}

\begin{figure*}
\centering
\includegraphics[width=0.48\linewidth,trim=1cm 1cm 1cm 1cm,clip]{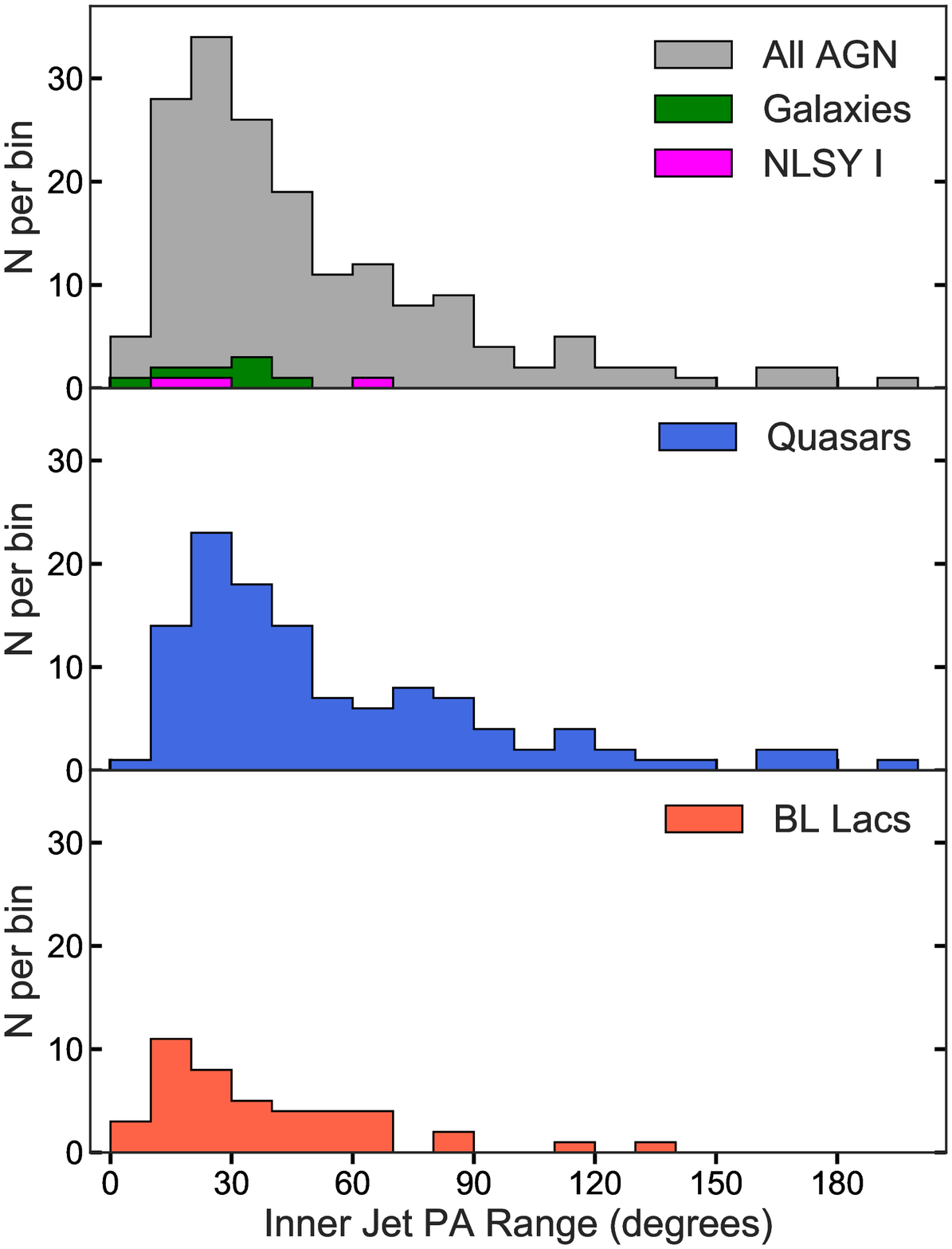}
\includegraphics[width=0.48\linewidth,trim=1cm 1cm 1cm 1cm,clip]{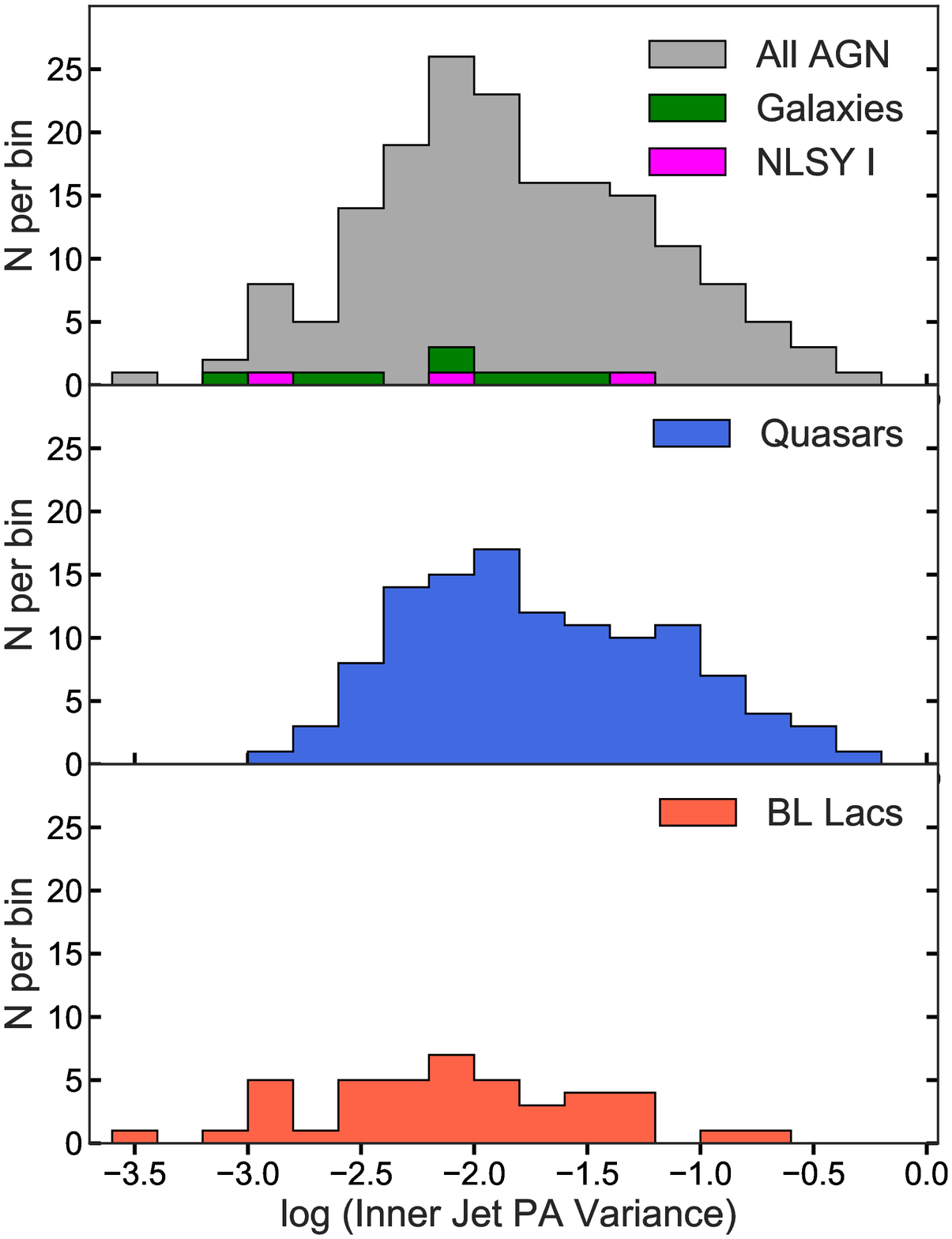}
\caption{\label{PAHIST_OPTCLASS} Distributions of inner
  jet PA range (left panel) and variance (right panel) for
  different AGN optical classes.}
\end{figure*}

\subsubsection{Jet Nozzle Wobbling}

There has been extensive discussion in the literature about possible
causes of nozzle wobbling in AGN jets (see, e.g., review by
\citealt{2014RAA....14..249Q}), which include MHD flow instabilities
\citep{2015AstL...41..712M}, Lense-Thirring precession of the
accretion disk \citep{2002ApJ...573L..23L, 2004ApJ...616L..99C}, or
the influence of a binary black hole companion
\citep{2021MNRAS.503.4400D}.  Precession is well-characterized in
stellar jet systems such as SS 433 \citep{2008ApJ...676..584R}, but
the much longer timescales and vastly larger distances associated with AGN jets and their accretion disks make their study more challenging.

A complicating factor in using VLBI data to investigate jet precession
is that we are not seeing the full extent of the jet. Our long term
monitoring has shown that at any given epoch, only certain portions of
the flow are sufficiently energized to appear in VLBA images, which
typically have dynamic ranges of a few thousand to one
\citep{MOJAVE_X}. A good example is the quasar 1308+326
(\objectname{OP 313}), which in the mid 1990s showed no apparent
downstream jet emission on parsec scales \citep{2004ApJS..150..187O}.
Throughout the following 25 years this AGN launched a series of bright
features that moved downstream on trajectories with a variety of
individual position angles, giving an illusory bent jet appearance in
individual epoch images. When the images over a 25 year period are
stacked, however, a smooth conical jet structure emerges
\citep{MOJAVE_XIV}.  In the case of 1308+326, the
axis of the broader outflow has remained relatively stable, but the
inner jet PA has varied by $\sim 85^\circ$ due to individual
emerging features that do not fill the entire jet cross-section.

To look for possible evidence  of nozzle wobbling, we examined 24 jets
in our sample that display overall monotonic trends of 10 years or
more in their PA evolution. In 11 cases, the evolution is
caused by one or more jet features moving on non-radial or curved
trajectories. In the remaining 13 AGN, however, multiple features
emerge from the core over time, yet there is an overall monotonic evolution of the jet PA. 
In other words, the ejection directions of successive
features are not random, but follow a systematic trend.
One example is 0430+052 (\objectname{3C 120}), for which we have sufficient spatial
resolution ($z = 0.033$; 0.65 pc per mas) to track the motions of many
features that are ejected at a rate of roughly one per year. Over a 25
year period, the initial PAs of successive features have progressed steadily
from $\sim 231^\circ$ to $\sim 256^\circ$. The bright nearby quasar
1226+023 (\objectname{3C 273}; $z = 0.1583$) has also displayed an overall swing in ejection PAs
from $\sim 237^\circ$ in 1995 to $\sim 208^\circ$ in 2008. After 2008,
its inner jet PA remained relatively stable at $\sim 220^\circ\pm
10^\circ$ (\autoref{PAvstime2}). 

Because the ejected features do not fill the entire jet
cross-section in observed emission, a wobbling flow instability generated near the jet base
is the most likely mechanism for the observed PA variations in many
jets. As this region of enhanced magnetic field strength and/or plasma
density precesses, new jet features are launched at different
locations within a broader outflow of typical full opening angle $\sim 1^\circ -4^\circ$ (Section \ref{jetPAclasses}).  Occasionally this instability may
be disrupted and/or a new instability forms at another location,
resulting in a sudden jump in jet PA, as seen in \objectname{OJ 287}. 
\cite{2021MNRAS.504.3823W} describe a similar scenario for the nearby radio-quiet quasar Mrk 231, in which they report a $\sim 60^\circ$ change in pc-scale jet position angle over a 25 year period. In their model the jet occupies a wide angle cone, and the bright knots represent the active working surface of the jet head hitting the interstellar medium. 

Several jets in our sample have very large PA ranges ($>150^\circ$):
\objectname{B2 0202+319}, 0355+508 (\objectname{NRAO 150}),
\objectname{S4 1144+402}, \objectname{B3 1417+385}, \objectname{PKS
  1622$-$253}, 2028+492 (\objectname{87GB 202807.5+491605}), which may be indicative of
a viewing angle inside the jet opening angle. Since the
individual moving features do not fill the entire jet cross section,
such jets would be expected to display a wide range of ejection position angles.

If the main axis of the broad outflow remains stable over time, a wobbling
jet instability could appear as an oscillatory PA trend in our data, provided that the precession period is smaller than our time coverage (up to 25 years
for the most heavily observed jets).  In \cite{MOJAVE_X} we reported
12 jets with possible oscillating PAs, but the fitted periods
of 5 yr to 12 yr were too long to establish any periodicity.

Of the 173 jets in our sub-sample with good epoch coverage, 67 show
possible back and forth PA evolution.  We used the Lomb-Scargle
periodogram algorithm for unequally sampled data
\citep{1976ApSS..39..447L,1982ApJ...263..835S} to look for possible
periodicities. We plot the best fit sinusoid cases in
\autoref{sinefits}.  The jet with the lowest reduced $\chi^2$ fit
value is 0738+313 (\objectname{OI 363}), but the fitted period of 16.7
yr is comparable to the 24 yr range of the data. The other fits in
\autoref{sinefits} have larger $\chi^2$ values due to departures from
pure sinusoidal periodic behavior.  The best fit periods range from 6 yr to 16.7 yr.  Given that these are comparable to the length of our VLBA
coverage, it is not possible with our current data set to robustly claim PA periodicity in any of
the jets in our sample.  We note that 8 of the 12 oscillating PA
jets we reported in \cite{MOJAVE_X}: \objectname{PKS 0754+100},
1308+326, \objectname{PKS 1335$-$127}, 1611+343 (\objectname{DA 406}),
1730$-$130 (\objectname{NRAO 530}), \objectname{S5 1803+784},
\objectname{PKS 2134+004}, 2145+067 (\objectname{4C +06.69}) do not
show sinusoidal periodicity, based on our new PA determinations and/or the addition of new VLBA epochs.

\begin{figure*}
\centering
\includegraphics[width=0.48\linewidth,trim=1cm 0.1cm 1cm 1cm,clip]{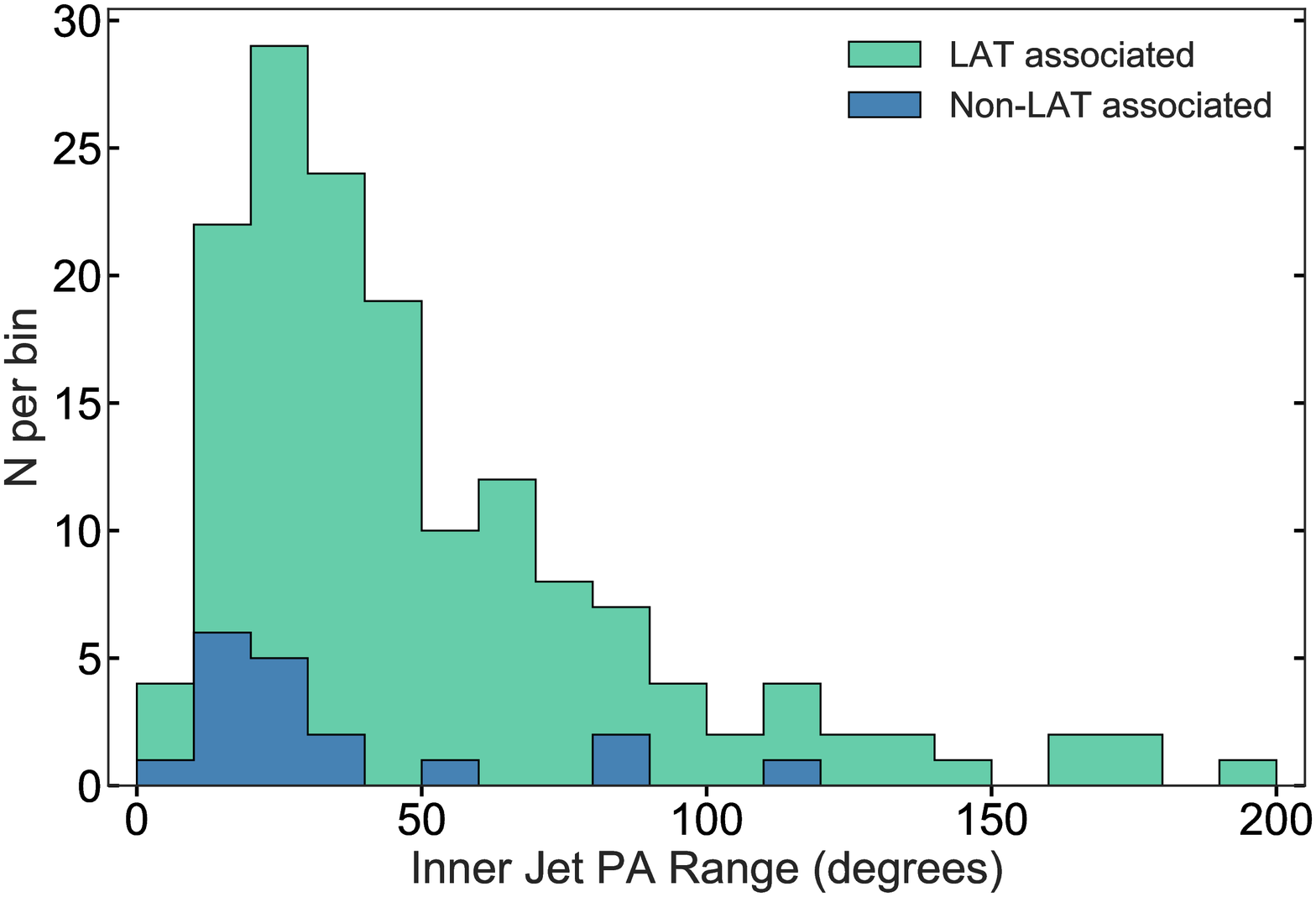}
\includegraphics[width=0.48\linewidth,trim=1cm 0.1cm 1cm 1cm,clip]{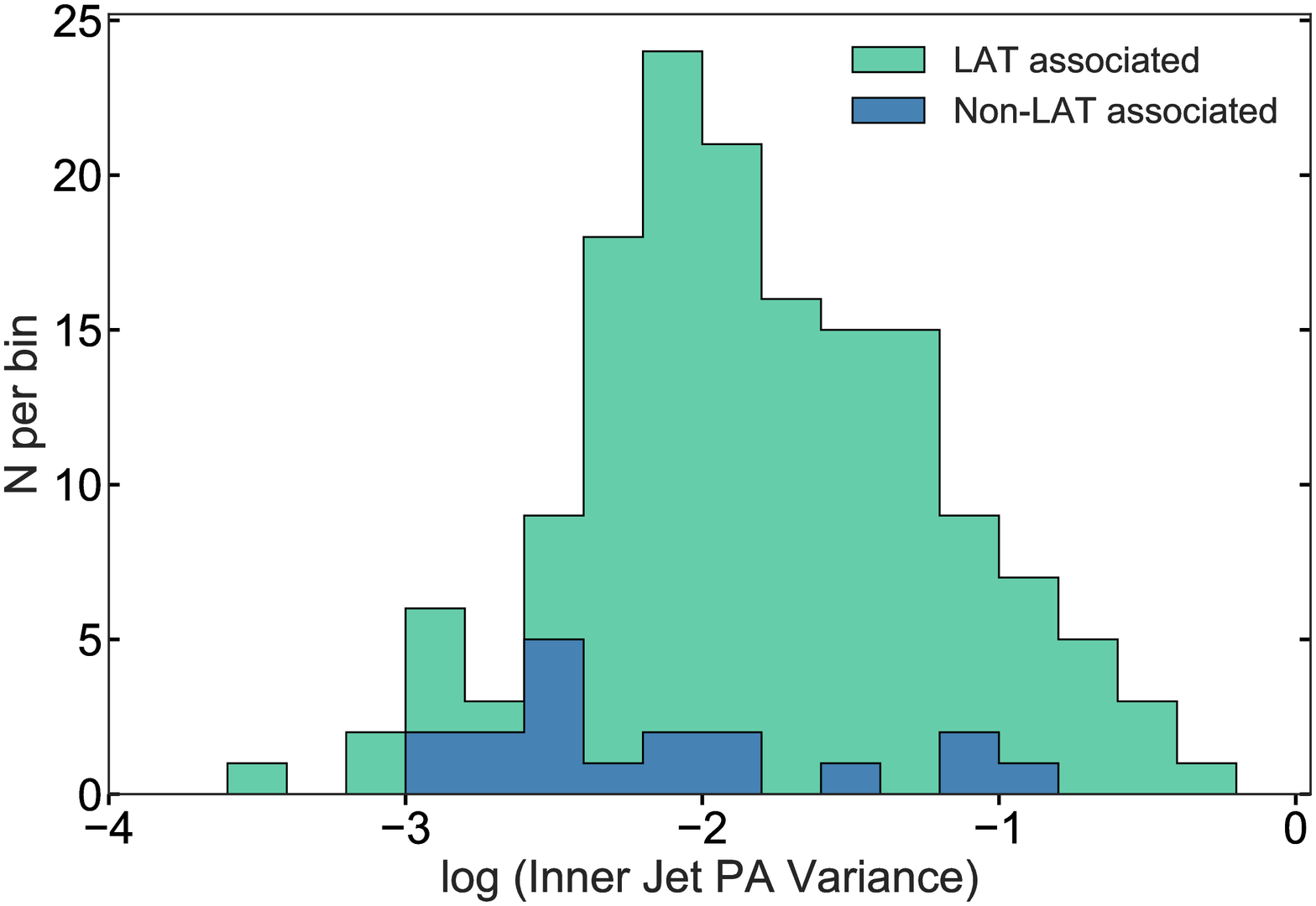}
\caption{\label{PAHIST_LATCLASS} Distributions of inner
  jet PA range (left panel) and variance (right panel) for
  LAT associated and non-LAT associated AGN.}
\end{figure*}

\subsubsection{\label{jetPAclasses} Trends with AGN Classes}

We have examined the PA range and variance statistics with respect to
general AGN properties using Anderson-Darling tests 
\citep{Stephens1974}.  Our sub-sample is made up of 117 quasars, 43
BL~Lacs, 10 radio galaxies, and 3 narrow-line Seyfert Is. All but 18
of these AGN have been listed as {\it Fermi} LAT gamma-ray
associations (\autoref{gentable}).

With respect to optical classification, we find that BL~Lacs have
smaller PA ranges ($p_\mathrm{null} = 0.0037$) and PA variances
($p_\mathrm{null} = 0.001$) than quasars (\autoref{PAHIST_OPTCLASS}).
Our tests indicate that these two sub-samples have indistinguishable
VLBA epoch coverage, but substantially different redshift
distributions, with the BL~Lacs having generally lower $z$ values. It
is unlikely that the PA differences are due to redshift, however, since
we find (i) no trend of $\Delta$PA with $z$ within the quasar
sub-sample (which spans $0.16 < z < 3.4$) and (ii) no significant difference in the $\Delta$PA distributions of quasars and radio galaxies, where
the latter are all at low redshift ($0.004 < z < 0.14$).

We do not find any statistically significant differences in the PA
ranges or PA variances of the 23 TeV-detected AGN compared to the
other AGN in our sample. However, we find that the 18 non-LAT
associated AGN have smaller PA ranges ($p_\mathrm{null} = 0.0098$) and PA
variances ($p_\mathrm{null} = 0.0076$) than the LAT associated AGN
(\autoref{PAHIST_LATCLASS}). There are no significant differences in the epoch
coverage or redshift distributions of the LAT vs. non-LAT AGN. 

There are several possible factors that could increase the range and
variance of a jet's PA over time, including a wider apparent opening
angle of its overall plasma outflow, and a less stable jet base.
Also, as more emerging features are tracked in a jet over time, the
spread of initial trajectory position angles ($\Delta$ PA) may be
expected to increase until it approaches the full jet opening angle.
The BL~Lacs, quasars, LAT and non-LAT AGN show no significant
differences in their number of robust jet features, which suggests
that the feature ejection rate (which is not affected by Doppler time
compression effects) does not have a strong influence on the PA
variance. We have also ruled out the possible effects of observational
time coverage and redshift based on our statistical tests.  The
remaining factors to consider are the intrinsic nozzle stability,
intrinsic opening angles, and viewing angles of the jets.

In a previous analysis of the MOJAVE sample \citep{2015ApJ...810L...9L}
we found that the non-{\it Fermi} LAT associated AGN tend to have lower Doppler boosting factors and spectral energy distributions (SED) that are
peaked at lower frequencies. All of the 43 BL~Lacs in our sub-sample
are LAT-associated, which is consistent with their generally harder
gamma-ray spectra and SED peaks that span a large range, up to
$10^{17}$ Hz. The non-LAT radio galaxies and quasars, on the other
hand, have SED peaks that range up to only $10^{13.4}$ Hz.

\begin{figure}
\centering
\includegraphics[width=\linewidth,trim=1cm 0.5cm 2cm 1.5cm,clip]{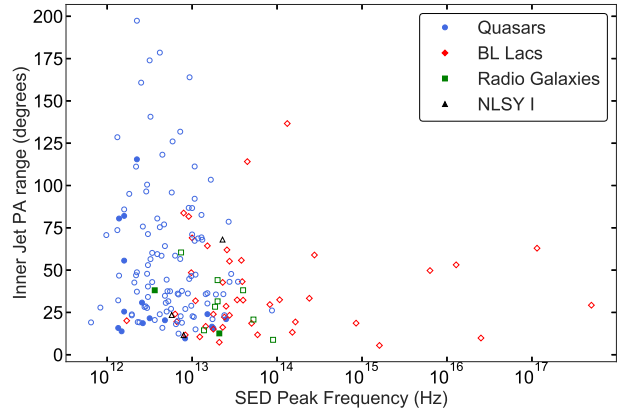}
\caption{\label{PArangevsSEDpeak} Scatter plot of $\Delta$PA versus SED peak
  frequency for the 174 AGN in our statistical sub-sample. The filled
  symbols indicate non-LAT associated AGN. }
\end{figure}

\begin{figure}
\centering
\includegraphics[width=\linewidth,trim=1cm 0.5cm 2cm 1.5cm,clip]{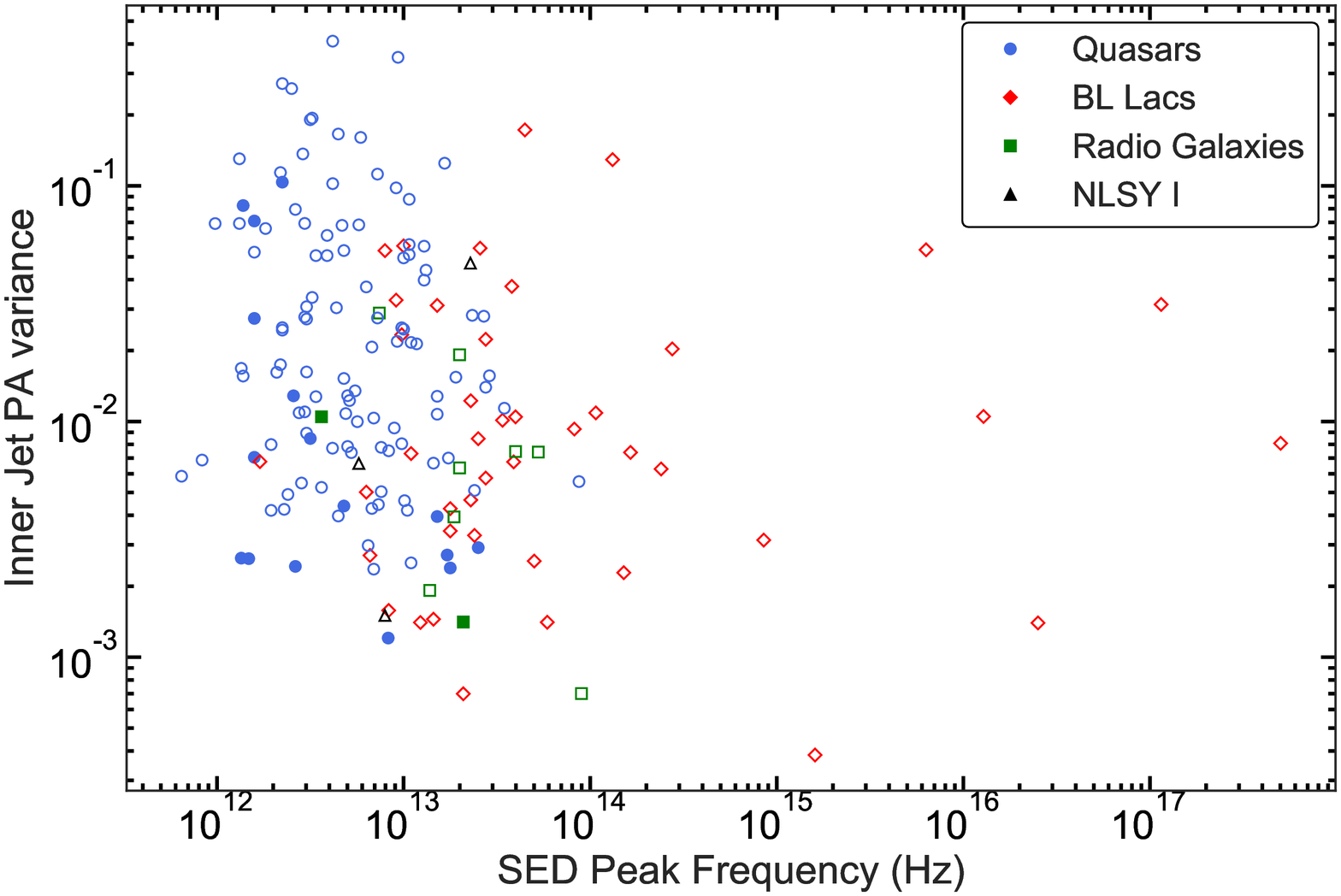}
\caption{\label{varPAvsSEDpeak} Scatter plot of PA variance versus SED peak
  frequency for the 174 AGN in our statistical sub-sample. The filled
  symbols indicate non-LAT associated AGN. }
\end{figure}

In \autoref{PArangevsSEDpeak} and \ref{varPAvsSEDpeak} we plot
Var(PA) and $\Delta$PA versus the SED peak frequency for the 174 AGN
in our sub-sample.  The non-LAT associated AGN are plotted with filled
symbols.  There is a tendency for lower synchrotron peaked AGN to have
more variable inner jet PAs ($p_\mathrm{null} = 0.006$ for $\Delta$PA and $p_\mathrm{null} = 0.0003$ for Var(PA), according to Spearman's $\rho$ tests).

We have determined Doppler factors ($\delta$) for the jets in our sample as part of a MOJAVE study of core brightness temperatures (Homan et al., in prep.).  The latter follows on our previous work \citep{2006ApJ...642L.115H} and exploits the fact that in their median low-brightness temperature ($T_b$) state, the MOJAVE AGN cores have a narrow range of intrinsic $T_b$ values ($T_\mathrm{int}$). Homan et al. derive Doppler factors from the observed $T_b$ values according to $T_\mathrm{obs} = \delta T_\mathrm{int}$, and jet viewing angles ($\theta$) from the maximum measured jet speeds presented in this paper. 

\begin{figure}
\centering
\includegraphics[width=\linewidth,trim=1cm 0.5cm 2cm 1.5cm,clip]{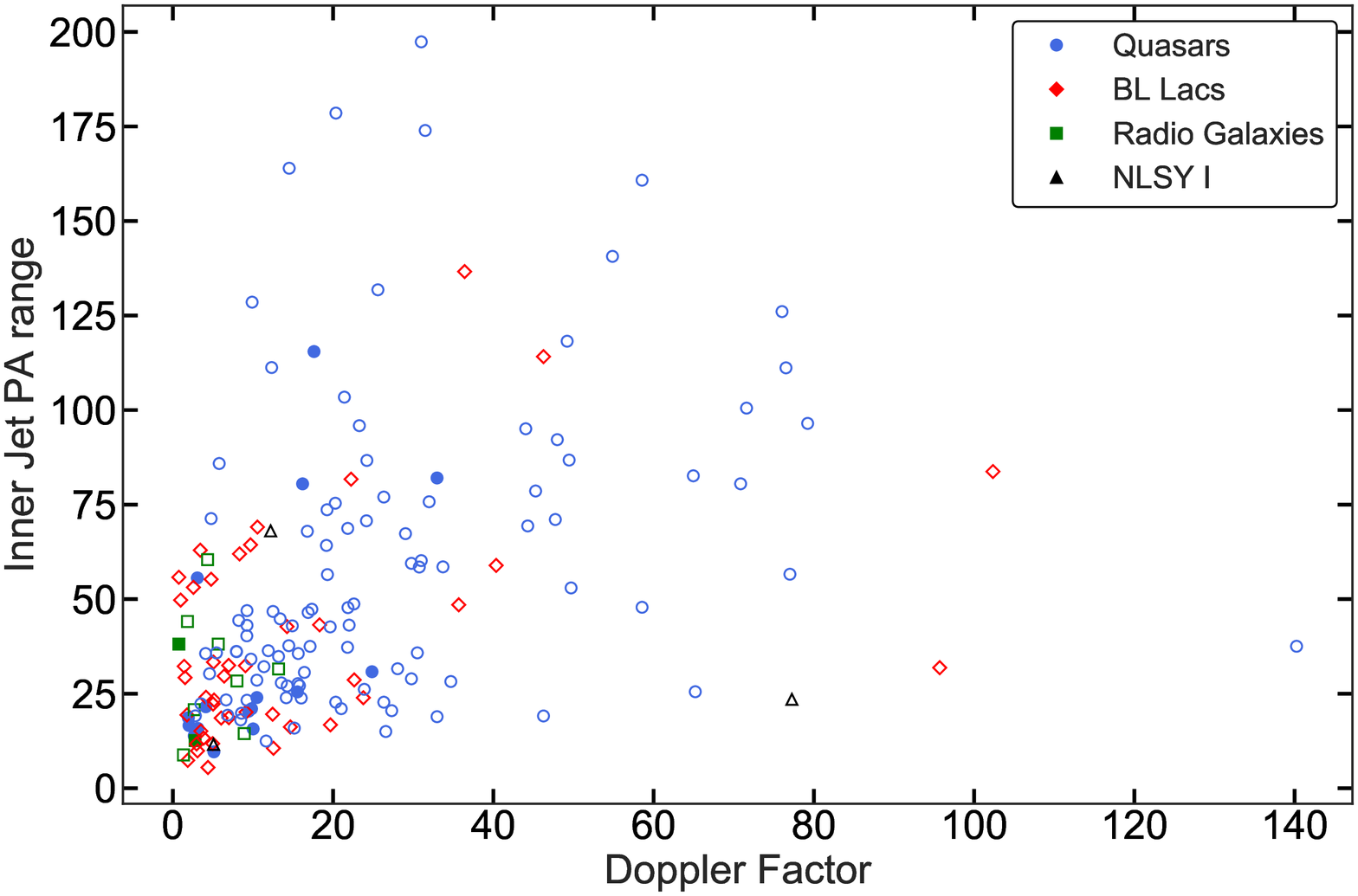}
\caption{\label{PArangevsdoppler} Scatter plot of inner jet PA range versus Doppler factor for different optical classes, with quasars = blue circles, BL~Lacs = red diamonds, radio galaxies = green squares, NLSY1 = black triangles.  Filled symbols indicate non-LAT associated AGN.  }
\end{figure}

\begin{figure}
\centering
\includegraphics[width=\linewidth,trim=1cm 0.5cm 2cm 1.5cm,clip]{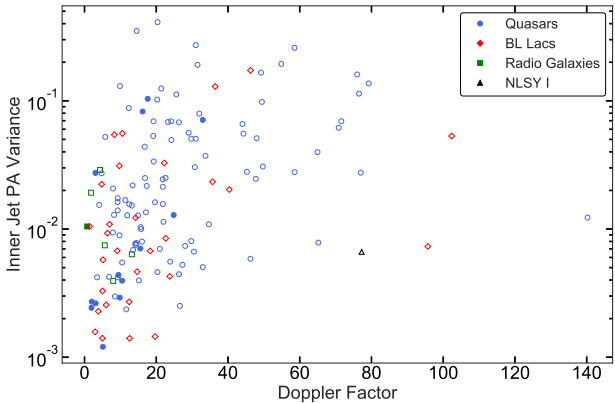}
\caption{\label{varPAvsdoppler} Scatter plot of PA variance versus Doppler factor for different optical classes, with quasars = blue circles, BL~Lacs = red diamonds, radio galaxies = green squares, NLSY1 = black triangles.  Filled symbols indicate non-LAT associated AGN.  }
\end{figure}

The jets in our sample with $\delta \lesssim 10$ show a smaller range of jet PA (\autoref{PArangevsdoppler}) and PA variance (\autoref{varPAvsdoppler}). This is primarily a consequence of their larger jet viewing angles. As a conical jet is oriented closer to the plane of the sky ($\theta \rightarrow 90^\circ$), its apparent half opening angle $\phi_\mathrm{app}$ will approach its intrinsic half opening angle $\phi$, where
\begin{equation}
\tan{\phi_\mathrm{app}} = {\tan{\phi} \over \sin{\theta}} \left(1 - {\tan^2{\phi} \over \tan^2{\theta}}\right)^{-1/2}.
\end{equation}
Conversely, as a conical jet is oriented closer to the line of sight, it will be viewed inside its opening angle when $\theta \le \phi$.

\begin{figure}
\centering
\includegraphics[width=\linewidth,trim=1cm 0.5cm 2cm 1.5cm,clip]{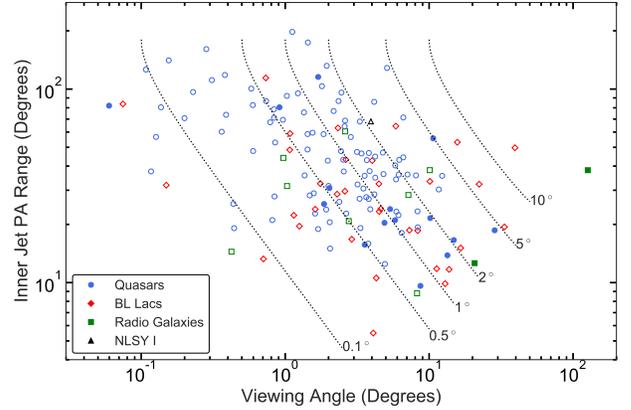}
\caption{\label{PArangevstheta} Scatter plot of inner jet PA range versus viewing angle for different optical classes, with quasars = blue circles, BL~Lacs = red diamonds, radio galaxies = green squares, NLSY1 = black triangles.  Filled symbols indicate non-LAT associated AGN.  The dotted curves indicate the apparent full opening angle of a conical jet as a function of viewing angle, for different values of intrinsic half opening angle as indicated. }
\end{figure}

In \autoref{PArangevstheta} we plot the inner jet PA range versus jet viewing angle. A Spearman non-parametric test indicates a strong trend ($p_\mathrm{null} = 10^{-10}$) of smaller PA ranges for jets viewed farther from the line of sight. We have over-plotted dashed curves showing the full apparent jet opening angle as a function of viewing angle $\theta$ for several values of intrinsic half opening angle. We have truncated the top of each plotted curve at $\theta = \phi$, beyond which the expected apparent PA range increases to $360\arcdeg$. As evident from \autoref{PArangevstheta}, the majority of the jets have intrinsic half opening angles between $\sim 0.5^\circ$ and $\sim 2^\circ$. 

It can be seen in \autoref{PArangevstheta} that there are no obvious
differences in the intrinsic jet opening angles of the BL~Lacs versus
quasars. The BL~Lacs in the MOJAVE sample are a mixture of low- and
high-spectral peaked blazars, with the latter generally having lower
Lorentz factors, lower Doppler factors, and larger viewing angles than
the quasars (Homan et al., ApJ, submitted). We conclude that the
overall larger viewing angles of BL~Lac jets are responsible for their
smaller jet PA ranges and variances.

The LAT and non-LAT associated AGN have a similar intrinsic median opening angle, although we note there is only one non-LAT jet (PKS 0607$-$15) narrower than $0.4^\circ$.  The non-LAT jets have larger viewing angles ($p_\mathrm{null} = 0.002$) and lower Doppler factors ($p_\mathrm{null} = 0.003$). Their respective Lorentz factor distributions have a marginal probability ($p_\mathrm{null} = 0.04$) of being drawn from the same population. As it was in the case of the BL~Lacs versus quasars, the non-LAT jets have smaller PA ranges and variances than LAT-associated AGN because they are oriented at larger angles to the line of sight.

\section{\label{conclusions}SUMMARY AND CONCLUSIONS}

We have analyzed the parsec-scale jet kinematics of 447 bright
radio-loud AGN based on 15 GHz VLBA data obtained between 1994 August
31 and 2019 August 4.  This represents the largest and most complete AGN jet kinematics study to date. These northern sky AGN (J2000 declination
$>-30\arcdeg$) have been part of the 2cm VLBA survey or MOJAVE
programs, and have correlated flux density $> 50$ mJy at 15 GHz.
There are 49 AGN that have not appeared in previous MOJAVE kinematics
papers. These were added on the basis of their gamma-ray emission
properties or membership in the Robopol optical polarization
monitoring program. We present new total intensity and linear
polarization maps obtained between 2017 January 1 to 2019 August 4 for
143 AGN in our sample.

By modeling the jet emission with a series of Gaussians in the
interferometric visibility plane, we identified and tracked 1923
individual features over at least five epochs. We fitted their sky
trajectories with simple radial and vector motion models, and
additionally carried out constant acceleration fits for 926 features
that had ten or more epochs.
\par
We  summarize our findings as follows:

1. We tracked at least one robust bright jet feature across five or
more epochs in 419 of the 447 AGN jets in our analysis. A majority
(60\%) of the well-sampled jet features showed evidence of either
accelerated or non-radial motion at the $\ge 3 \sigma$ level.

2. Only 2.5\% of the robust jet features had velocity vectors
apparently directed inward toward the core. However, 8.7\% of the AGN
jets in the sample had at least one inward-moving feature. These
features may be the result of centroid shifts in diffuse emission
regions, or curved trajectories crossing the line of sight.

3. We identified 64 non-accelerating features with unusually slow
pattern speeds ($\mu < 20$ \muasyr and at least 10 times slower than
the fastest feature in the jet) in 47 jets that may be standing shocks
in the flow.

4. We have analyzed variations of the innermost jet position angles in
our AGN sample over time. We restricted our analysis to 173 jets
that had 12 or more VLBA epochs over a minimum 10 year period. By
using the PA of the closest fitted Gaussian feature to the core at
each epoch, we derived a mean jet PA, as well as the range of jet PA
and its circular variance.  Most of the jets have inner PAs that vary
on decadal timescales over a range of $10^\circ$ to $50^\circ$.
However some jets display very large changes in PA, with the overall
distribution having a tail out to $\Delta{PA} = 200^\circ$. Some jets
show a monotonic evolution of PA after ejecting a feature that moves
outward on a curved or non-radial trajectory.  In some cases, a jet
experiences a sudden large jump in PA when a new bright feature
emerges with a substantially different trajectory than previously
ejected features.

5. We find that AGN with SEDs peaked at lower frequencies tend to have
more variable jet PAs. This is reflected in a tendency for the BL~Lacs in
our sample to have less variable PAs, since their SED peak distribution extends to much higher frequencies than the quasars and radio galaxies. Furthermore, the \fermi LAT gamma-ray associated AGN in our sample tend to have more variable PAs than the non-LAT AGN.  We have ruled out the possible effects of redshift, time sampling, and feature ejection rate as the cause of these differences. By using Lorentz factor and Doppler factor measurements from a MOJAVE analysis of AGN core brightness temperatures (Homan et al., ApJ, submitted), we conclude that the non-LAT and higher synchrotron peaked (BL~Lac) jets show smaller variance and range in their inner jet PAs because they are viewed at slightly larger angles to the line of sight. 
  
6. We have identified 13 AGN where over a decade long period, multiple features emerge from the core with ejection PAs that follow a systematic
trend. Since the ejected features do not fill the entire jet
cross-section, this behavior is indicative of a wobbling flow
instability near the jet base. New features may emerge from this region of
enhanced magnetic field/plasma density as it precesses on
$\sim$decadal timescales.  Occasionally this instability may
be disrupted and/or a new instability forms at another location,
resulting in a sudden jump in the inner jet PA. We have looked for evidence of periodic PA behavior in 67 jets that show back and forth PA evolution using
Lomb-Scargle periodograms. The best fit periods range from 6 y to 16.7
y, however, we cannot claim any bona fide cases of periodicity since
these periods are comparable to the 10 y -- 25 y VLBA time coverages
of the AGN.

\begin{acknowledgments}

We thank the anonymous referee for helpful comments that improved the manuscript. 
The MOJAVE project was supported by NASA-{\it Fermi} grants 80NSSC19K1579, NNX15AU76G and NNX12A087G.
YYK is supported in the framework of the State project ``Science'' by the Ministry of Science and Higher Education of the Russian Federation under the contract 075-15-2020-778.
%
TS was supported by the Academy of Finland projects 274477 and 315721. 
%
%
The Very Long Baseline Array and the National Radio Astronomy Observatory are facilities of the National Science Foundation operated under cooperative agreement by Associated Universities, Inc. 
This work made use of the Swinburne University of Technology software correlator \citep{2011PASP..123..275D}, developed as part of the Australian Major National Research Facilities Programme and operated under licence.
This research has used observations with RATAN-600 of the Special Astrophysical Observatory, Russian Academy of Sciences (SAO RAS). The observations with the SAO RAS telescopes are supported by the Ministry of Science and Higher Education of the Russian Federation.
This research has made use of data from the OVRO 40-m monitoring program \cite{2011ApJS..194...29R}, which is supported in part by NASA grants NNX08AW31G, NNX11A043G, and NNX14AQ89G and NSF grants AST-0808050 and AST-1109911. 
This research has made use of data from the University of Michigan Radio Astronomy Observatory which has been supported by the University of Michigan and by a series of grants from the National Science Foundation, most recently AST-0607523.
This research has made use of the NASA/IPAC Extragalactic Database (NED) which is operated by the Jet Propulsion Laboratory, California Institute of Technology, under contract with the National Aeronautics and Space Administration.

\end{acknowledgments}

\facilities{VLBA, OVRO:40m, RATAN, UMRAO, NED, ADS}

\software{astropy \citep{2013AA...558A..33A,2018AJ....156..123A}, DIFMAP
  \citep{DIFMAP}, AIPS \citep{AIPS}}

\appendix

\section{Notes on Individual AGN}
\label{s:source_notes}

Here we provide comments on individual AGN supplementing those given
in \cite{MOJAVE_X}, \cite{MOJAVE_XIII}, and \cite{MOJAVE_XVII}. 

0106+678 (\objectname{4C +67.04}): An additional VLBA epoch in 2019 has shown that a
jet feature (id  4) in this BL~Lac no longer has statistically
significant inward motion. 

0420+417 (\objectname{4C +41.11}): With the addition of new epochs after 2017.0, the
centroid of the large diffuse feature (id  6) at the end of the jet
shows inward motion. The evolution of the innermost feature (id  11)
is not consistent with radial, outward motion.

0518+211 (\objectname{RGB J0521+212}): Additional epochs obtained since
2017.0 indicate that the innermost jet feature in this BL
Lac object (id  10) no longer has statistically significant inward
motion in the vector fit, but continues to be inward in the acceleration fit. 

\objectname{S4 0636+680}: This quasar at  $z= 3.177$ has a radio
spectrum peaked at 5 GHz, and compact radio structure $\sim 1$ mas
in extent. We assume the core lies in the northernmost feature, which
has the highest brightness temperature and flux density. 

\objectname{PKS B0742+103}: This quasar has a radio spectrum peaked at 3
GHz, and an uncertain core location in our 15 GHz images. We
therefore classified none of the jet features as robust. We assigned
the core to the brightest jet feature at each epoch, which lies in
between two features (id  4 and id  5). A recent 43 GHz VLBA image
by \cite{2020ApJS..247...57C} shows a core-dominated morphology with
jet emission to the N and NW.

0743$-$006 (\objectname{OI $-$072}):  This quasar at $z = 0.996$ has a radio
spectrum peaked at 7 GHz and compact radio structure only 2 mas in
extent. A VLBA 43 GHz by \cite{2020ApJS..247...57C} shows the
brightest feature at the southern end of the jet, which we use as the
core location for our kinematics analysis. 

0810+646 (\objectname{87GB 081008.0+644032}): This BL~Lac at $z = 0.239$ had no bright robust jet
features.

\objectname{MRC 0910$-$208}: None of the jet features in this quasar
at $z=0.198$ were sufficiently bright or compact to be labeled as
robust. 

1101+384 (\objectname{Mrk 421}): Two features (id  9 and id  11) show
significant inward motion in the vector motion fit but not in the
acceleration fit. Another feature (id  8) has inward motion in both
fits. 

\objectname{PKS 1118$-$056}: New epochs obtained after 2017.0 show a new feature (id  8) to have inward motion. 

1142+198 (\objectname{3C 264}): Additional kinematics analyses of the VLBA data on
this AGN jet have been published by \cite{2020ApJ...896...41A} and
\cite{2019AA...627A..89B}.

1253$-$055 (\objectname{3C 279}): The VLBA epochs in 2013--2014
  and 2018 April 22 are affected by the emergence of very bright
  features ($> 10$ Jy). After these features were ejected they faded
  rapidly and moved downstream, passing through the quasi-stationary
  core. Thus the location of the 'true' jet core (which is normally
  too weak to be seen) was briefly revealed.  The reference 'core'
  position that we use in this jet at all epochs is likely a strong
  quasi-stationary feature in the downstream flow.  Our interpretation is based
  on the resulting consistent/continuous sky trajectories of features
  5,13, and 14 during 2013--2018 if the quasi-stationary feature is
  used as a reference point.

1509+054 (PMN J1511+0518): This radio galaxy has two-sided jet structure \citep{2006AA...450..959O}. We have assumed the eastern jet is the approaching one, based on its faster apparent speed (id 3). 

1652+398 (\objectname{Mrk 501}): One jet feature (id  5) has significant inward motion in the
acceleration fit, but not in the vector model fit. 
 
\objectname{PKS 1725+123}: The addition of new epochs obtained since 2017.0
establish inward motion in one feature (id  2).

1754+155 (\objectname{87GB 175437.6+153548}): We could not identify any robust jet
features in this quasar at $z = 2.06$.

1928+738 (\objectname{4C +73.18}): In 2012 an apparent counter-jet feature (id
17) with an apparent speed of 0.4 $c$ emerged in this quasar jet. This
was followed by another feature (id 21) that appeared in 2017 farther
upstream from the core, with no statistically significant motion ($20
\pm 20 \muasyr$). Given the fast superluminal speeds in the
approaching jet (8.4 $c$) and expected large de-boosting of the
receding jet emission, these observations suggest that these aren't
true counterjet features, and that the true core in this quasar is 
visible only at some epochs.

2013$-$092 (\objectname{PMN J2016$-$0903}): All of the jet features in this BL~Lac with
unknown redshift were too faint to be considered robust. 

2115+000 (\objectname{PMN J2118+0013}): The radio structure of this narrow-lined Sy
1 galaxy at $z = 0.463$ from the \fermi hard-spectrum sample was too
compact to identify any robust jet features.

2234+282 (\objectname{CTD 135}): A 43 GHz VLBA image by \cite{2020ApJS..247...57C}
indicates that the core is located at the extreme southwest end of the jet. 

\objectname{S5 2353+816}: No robust jet features could be identified in this BL~Lac
object due to the large gap in the VLBA time coverage.

\bibliography{lister}
\bibliographystyle{aasjournal}

\end{document}